\documentclass[aps,pre,reprint,superscriptaddress,amsmath,amssymb]{revtex4-2}
\usepackage{graphicx}
\usepackage{dcolumn}
\usepackage{bm}
\usepackage[mathlines]{lineno}
\usepackage{epsfig}
\usepackage{xcolor,colortbl}
\usepackage{tabularx}
\usepackage{epstopdf}
\usepackage{amsmath,amssymb,amsthm}
\usepackage{bm}
\usepackage{float}
\usepackage{xcolor}
\usepackage{mathtools}
\usepackage{subfig}
\usepackage{hhline}
\usepackage{algcompatible}
\usepackage{newfloat}
\usepackage[justification=RaggedRight]{caption}
\usepackage[bookmarks=true,colorlinks=false]{hyperref}
\DeclareFloatingEnvironment[
fileext=loa,
listname=List of Algorithms,
name=ALGORITHM,
placement=tbhp,
]{algorithm}

\begin{document}
	\title{Lagrangian features of turbulent transport in tokamak plasmas}
	\author{D. I. Palade}
	\email{dragos.palade@inflpr.ro}
	\affiliation{National Institute of Laser, Plasma and Radiation Physics,	M\u{a}gurele, Bucharest, Romania}
	\affiliation{Faculty of Physics, University of Bucharest, Măgurele, Romania}

\date{\today}

\begin{abstract}
This study investigates the Lagrangian properties of ion turbulent transport driven by drift-type turbulence in tokamak plasmas. Despite the compressible and inhomogeneous nature of Eulerian gyrocenter drifts, numerical simulations with the T3ST code reveal approximate ergodicity, stationarity, and time-symmetry. These characteristics are attributed to broad initial phase-space distributions that support ergodic mixing. Moreover, relatively minor constraints on the initial distributions are found to have negligible effects on transport levels.
\end{abstract}


\maketitle

\section{Introduction}
\label{Section_1}

Turbulent flows are ubiquitous in nature, filling the gap between smooth (deterministic) and chaotic (random) motion. Their Eulerian properties, described by the velocity vector field $\mathbf{v}(\mathbf{x},t)$ and its space-time statistics, are often easy to measure, simulate and interpret. In contrast, Lagrangian characteristics are significantly more challenging to capture and have been the subject of an extensive body of research over the past century being described by the statistics of particle trajectories $\mathbf{x}(t|\mathbf{x}_0)$ and associated Lagrangian velocities $\mathbf{v}^L(t|\mathbf{x}_0)\equiv \mathbf{v}(\mathbf{x}(t|\mathbf{x}_0), t)$. The relation between these two is the V-Langevin equation \cite{BALESCU200762}:

\begin{eqnarray}
	\label{eq_1.1}
\frac{d\mathbf{x}(t|\mathbf{x}_0)}{dt} = \mathbf{v}(\mathbf{x}(t|\mathbf{x}_0), t) \equiv \mathbf{v}^L(t), \quad \mathbf{x}(0|\mathbf{x}_0) = \mathbf{x}_0.
\end{eqnarray}

At the level of transport, it has long been known—since the seminal work of G.~I.~Taylor \cite{taylor1922diffusion}—that turbulent flows can drive diffusive transport, in a similar way the white noise involved in Brownian motion does. Building on Taylor’s pioneering work, Monin, Yaglom, and Lumley \cite{monin1971statistical} revealed important symmetry properties of Lagrangian quantities. If the Eulerian velocity field $\mathbf{v}(\mathbf{x}, t)$ is homogeneous in space, stationary in time, and divergence-free $\nabla\cdot\mathbf{v}\left(\mathbf{x},t\right)=0$, then the Lagrangian velocities $\mathbf{v}^L(t|\mathbf{x}_0) \equiv \mathbf{v}(\mathbf{x}(t), t)$ are also stationary. This implies that the distribution of Lagrangian velocities is invariant in time and identical to the distribution of Eulerian velocities, i.e., $P[\mathbf{v}^L(t|\mathbf{x}_0)] = P[\mathbf{v}(\mathbf{x}_0, t)]$ if the initial conditions $\mathbf{x}_0$ are distributed across the entire physical space. Furthermore, Lagrangian stationarity manifests in two-point statistics, as captured by the velocity autocorrelation function:

\begin{eqnarray}
	\label{eq_1.2}
\hat{L}(t, t') = \langle \mathbf{v}^L(t) \otimes \mathbf{v}^L(t') \rangle - \langle \mathbf{v}^L(t) \rangle \otimes \langle \mathbf{v}^L(t') \rangle,
\end{eqnarray}
which obeys $\hat{L}(t, t') \equiv \hat{L}(|t - t'|)$. Stationarity is typically accompanied by time-symmetry.

The Eulerian and Lagrangian perspectives meet inevitably in the definition of transport coefficients. A passive scalar $n(\mathbf{x},t)$ advected by the velocity field $\mathbf{v}(\mathbf{x},t)$ experiences a mesoscopic flux $\mathbf{\Gamma}$ that, if the dynamics is local, can be expanded in the spirit of Fick's law $\mathbf{\Gamma} = \mathbf{V} n -\hat{D}\nabla n$. The transport coefficients $\mathbf{V}, \hat{D}$ are interpreted as convection and diffusion and can be related \cite{taylor1922diffusion} to Lagrangian quantities (particle trajectories) as:
\begin{eqnarray}\label{eq_1.3a}
	\mathbf{V}(t) &=& \frac{d}{dt}\langle \mathbf{x}(t|\mathbf{x}_0)\rangle = \langle \mathbf{v}^L(t|\mathbf{x}_0)\rangle,\\	
	\hat{D}(t)&=& \frac{1}{2}\frac{d}{dt}\left(\langle \mathbf{x}^2(t|\mathbf{x}_0)\rangle - \langle \mathbf{x}(t|\mathbf{x}_0) \rangle^2 \right) =\label{eq_1.3b}\\
	& & = \int_0^t \hat{L}(t, \tau) \, d\tau = \int_0^t \hat{L}(\tau) \, d\tau.
\end{eqnarray}
where by $\langle\cdot\rangle$ we understand space-average over the distribution of initial conditions $\mathbf{x}_0\in\Omega$. 

Although in reality, in a single experiment, there is a single turbulent field $\mathbf{v}(\mathbf{x},t)$, it is possible to use a statistical description of turbulence \cite{balescu2005aspects,BALESCU200762,orszag1973statistical} starting from the following observation: 

\begin{quote}
\emph{For particles dispersed in a homogeneous, stationary turbulent field, their motion can often be treated as approximately ergodic. Consequently, ensemble-averaged statistics over multiple realizations of the field are representative of space-time averages within a single realization, allowing particle dynamics to be equivalently described either by many particles in one field or by a particle in an ensemble of statistically similar fields.}
\end{quote}

This is part of the \emph{ontic} argument behind the use of statistical ensembles in studies of turbulent transport. 
The other argument is \emph{epistemic} but, again, can't be rigorously motivated outside the assumption of ergodicity: the inherently chaotic nature of turbulence and its sensitivity to initial conditions makes it impossible to determine precisely its space-time configuration, thus, one is forced to resort to a statistical ensemble describing all possible states of turbulence akin to a \emph{canonical ensemble} from statistical thermodynamics where only probabilistic predictions can be made. 

While the statistical properties of turbulent transport have been extensively studied in general fluid and plasma contexts, their implications for magnetically confined fusion plasmas—such as those found in tokamaks—require special attention. In these systems, turbulence plays a decisive role in driving cross-field transport, undermining the very confinement these devices aim to achieve. Among the most promising approaches to fusion energy, tokamaks \cite{wesson2011tokamaks} use strong magnetic fields to trap high-temperature plasmas, but are fundamentally limited by drift-type turbulence that gives rise to radial transport. For these reasons, the modelling and prediction of turbulent transport is of paramount importance \cite{Mantica_2020}.

To a good approximation, the turbulent electrostatic potential $\phi(\mathbf{x},t)$ in a tokamak satisfies the assumptions of space-time stationarity. However, particle motion in these environments is influenced by a complex velocity field $\mathbf{v} = \mathbf{v}_{dr} + \mathbf{v}_{E \times B}$ (to be detailed in Section \ref{Section_2.1}), comprising a deterministic magnetic component ($\mathbf{v}_{dr}$ - the magnetic drifts) and a turbulent part $\mathbf{v}_{E \times B} = \mathbf{b}/B \times \nabla \phi$. Because the magnetic field $\mathbf{B}$ in tokamaks is inherently inhomogeneous, the resulting velocity field $\mathbf{v}$ is neither perfectly homogeneous, nor divergence-free.

The objective of the present work is to address, numerically, the following concerns:
\begin{quote}
	\emph{Given that the particle's drifts in tokamak devices are Eulerian-inhomogeneous and compressible, is the Lagrangian turbulent transport stationary, ergodic or time-reversible?}
\end{quote}

To the best of the author's knowledge, these questions have not been answered, by numerical means, before. The affirmative answer is often taken for granted—partly for convenience, and partly due to the observation that the deviations from Eulerian stationarity or compressibility tend to be relatively mild and relevant on mesoscopic space-scales. In this work, numerical investigations are carried out using the newly developed T3ST code \cite{palade2025t3st}. T3ST is a high-performance framework designed to track test-particle trajectories in tokamak environments under the combined influence of magnetic drifts and turbulent fields \cite{pomarjanschi2024neural,Palade_2024_scaling,Palade_2023}. It offers an ideal platform for analyzing Lagrangian quantities.

The remainder of this paper is organized as follows: Section~\ref{Section_2} reviews particle dynamics in tokamak environments and outlines the main features of the T3ST code. Section~\ref{Section_3} presents numerical answers to the questions posed above. Section~\ref{Section_4} concludes with a discussion and future perspectives.

\section{Theory and numerical simulations}
\label{Section_2}

\subsection{Description of turbulent transport in tokamaks}
\label{Section_2.1}

Viable, controlled thermonuclear fusion has not yet been achieved, despite more than seven decades of research \cite{Mantica_2020}. The most promising experimental approach remains magnetic confinement within tokamaks \cite{wesson2011tokamaks}, the latter being toroidal devices immersed in strong magnetic fields, designed to trap charged particles along closed trajectories.

Unfortunately, tokamaks (and other fusion devices, for that matter) suffer from a persistent level of radial transport, which is arguably one of the main obstacles to achieving effective confinement. This transport can be either (neo)classical—arising from particle collisions \cite{hinton1983collisional}—or anomalous, originating from turbulent electromagnetic fields \cite{balescu2005aspects}. In most relevant tokamak regimes turbulent dynamics dominates transport and for these reasons it is of paramount importance to achieve predictive modelling. 

Technically, turbulent transport in tokamaks is approximately local \cite{PhysRevE.82.025401}, meaning that the particle flux $\Gamma$ obeys a Fick's law of the form $\Gamma \approx V n-D\nabla n$, where $n$ is the particle density and $V,D$ are the convection and diffusion coefficients. The latter can be related to particle trajectories, as described in Section \ref{Section_1} via Eqs. \ref{eq_1.3a}-\ref{eq_1.3b} or, more precisely, as it will be discussed later, via Eqs. \ref{eq_2.1.4a}-\ref{eq_2.1.4b}.

The motion of charged particles in strongly magnetized plasmas is best described at the level of gyrocenters $(\mathbf{X}, v_\parallel, \mu)$, which replace the real phase space $(\mathbf{x}, \mathbf{u})$ by averaging out the fast and small-scale Larmor gyration \cite{Littlejohn}. We assume, embedded in a strong equilibrium magnetic field $\mathbf{B}$, with magnitude $B = |\mathbf{B}|$ and unit vector $\mathbf{b} = \mathbf{B}/B$, the presence of a turbulent electrostatic potential $\phi$, but neglect (for simplicity) collisions, plasma rotation, zonal flows, polarization drifts, and magnetic fluctuations. Under gyrokinetic ordering and Lie-perturbation theory \cite{Littlejohn}, the dynamics of the gyrocenters can be described by the following equations \cite{Brizard_hahm_gyrokinetic}:
\begin{eqnarray}
	\label{eq_2.1.1a}
	\frac{d\mathbf{X}}{dt} &=& v_\parallel \frac{\mathbf{B}^\star}{B_\parallel^\star} + \frac{\mathbf{E}^\star \times \mathbf{b}}{B_\parallel^\star},\\
	\label{eq_2.1.1b}
	\frac{d v_\parallel}{dt} &=& \frac{q}{m} \frac{\mathbf{E}^\star \cdot \mathbf{B}^\star}{B_\parallel^\star}.
\end{eqnarray}
The effective fields are given by $\mathbf{E}^\star = -\mu/q\nabla B -\nabla \phi^{gc}, \mathbf{B}^\star = \mathbf{B}+m/q v_\parallel\nabla\times\mathbf{b}$. While realistic magnetic fields typically have complex topologies best expressed in contravariant coordinates aligned with magnetic flux surfaces, we consider here a simplified circular equilibrium:
\begin{eqnarray}
	\label{eq_2.1.2}
	\mathbf{B} = B_0 R_0 \left( \nabla \varphi + \frac{r b_\theta(r)}{R} \nabla \theta \right).
\end{eqnarray}
$B_0$ is the field magnitude at the magnetic axis ($r=0$, $R=R_0$), and $b_\theta(r) = r/\bar{q}(r)/\sqrt{R_0^2 - r^2}$ characterizes the poloidal component. For the safety factor $\bar{q}(r)$, the analytical form $\bar{q}(r) = c_1 + c_2 r + c_3 r^2$ is used. The geometry follows the COCOS=2 convention \cite{SAUTER2013293} and is described in right-handed toroidal coordinates $(r, \theta, \varphi)$ with $R = R_0 + r \cos \theta$.

Breaking down the expressions from Eqs. \ref{eq_2.1.1a}-\ref{eq_2.1.1b} one can identify two types of components of motion. First we have the magnetic drifts $\mathbf{v}_{dr} \approx v_\parallel \mathbf{b}+\left(m v_\parallel^2+\mu B\right)/qB\nabla \ln B\times \mathbf{b}$ that emerge from the toroidally curved, large-scale magnetic field $\mathbf{B}$ and are fully deterministic \cite{etde_6135653}. Second, turbulent forces can be effectively reduced to the $\mathbf{E} \times \mathbf{B}$ drift, $\mathbf{v}_{E\times B} \approx -\nabla\phi^{gc} \times \mathbf{B}/B^2$, and the associated parallel acceleration, $a_\parallel \approx -q/m \nabla\phi^{gc} \cdot \mathbf{B}/B$. 

From now on we shall call the physical regime without turbulence (or collisions) \emph{quiescent}.

In these expressions, the electric potential $\phi$ characterizes the turbulent fluctuations and primarily originates from ion temperature gradient (ITG) or trapped electron mode (TEM) instabilities \cite{RevModPhys.71.735}. Other sources, such as electron temperature gradient (ETG) modes \cite{Jenko_ETG_10.1063/1.874014}, are generally not relevant to ion transport. Note however that the potential's derivatives are evaluated at the gyrocenter level, indicated by the superscript "gc". This introduces a finite Larmor radius (FLR) correction: $\tilde{\phi}^{gc}(\mathbf{k},t) = \tilde{\phi}(\mathbf{k},t)J_0(k_\perp\rho_L)$ where $\rho_L = mv_\perp/qB = \sqrt{2m\mu/q^2 B}$ is the Larmor radius, $k_\perp = |\mathbf{k}_\perp|$, and $\mathbf{k}_\perp = \mathbf{k} - k_\parallel \mathbf{b}$ with $k_\parallel = \mathbf{k} \cdot \mathbf{b}$.

Experimental evidence and gyrokinetic simulations suggest that, from an Eulerian perspective, $\phi(\mathbf{x}, t)$ is chaotic, microscopic (fluctuates on small time-space scales), almost normal distributed, time-stationary and space-homogeneous. The space homogeneity may not be present in the linear phase of drift instabilities when the ballooning structure dominates or at macroscopic scales where turbulence amplitude has variations, but it holds some validity in investigations of local transport in saturated turbulence. As discussed in the Introduction \ref{Section_1}, all these features enable one to employ statistical descriptions of turbulence in which the real turbulent field is replaced by a statistical ensemble of random fields that obey the space-time statistics of realistic turbulent realizations \cite{balescu2005aspects}. Within such ensembles, $n$-point distribution of potential values is Gaussian, the field has zero mean, $\langle \phi(\mathbf{x}, t) \rangle = 0$, and its statistical properties are encoded in the autocorrelation function
$E(\mathbf{x},t|\mathbf{x}^\prime,t^\prime) = \langle \phi(\mathbf{x},t) \phi(\mathbf{x}^\prime,t^\prime)\rangle\equiv E(\mathbf{x}-\mathbf{x}^\prime,|t-t^\prime|)$. Equivalently, the correlation can be written as Fourier transform of the turbulence spectrum $S(\mathbf{k}, \omega) = \langle|\tilde{\phi}(\mathbf{k,\omega})|^2\rangle$ \cite{Palade2021}.

Since the turbulence is of ITG origin, we employ a slab-inspired linear dispersion relation for the frequency:
\begin{eqnarray}
	\label{eq_2.1.3}
	\omega_\star(\mathbf{k}) = \frac{\mathbf{k} \cdot \mathbf{V}^\star_s}{1 + \rho_s^2 |k_\perp|^2},
\end{eqnarray}
where $\mathbf{V}^\star_s = -\nabla p \times \mathbf{b} / (n |e| B)$ denotes the ion diamagnetic velocity. This reflects in the frequencies of Fourier modes of turbulence which can be decomposed as $\omega = \omega_\star(\mathbf{k})+\Delta \omega$, where the statistics of $\Delta \omega$ is dictated through $S(\mathbf{k},\omega)$ and related intimately to the correlation time $\tau_c$. 

The transport coefficients can be connected to gyrocenter trajectories $\mathbf{X}(t|x)$ that evolve under the equations of motion \eqref{eq_2.1.1a}–\eqref{eq_2.1.1b} via convenient expressions:
\begin{eqnarray}
	\label{eq_2.1.4a}
	D(t | x) &=& \frac{1}{2} \frac{d}{dt} \left( \{ \langle \mathbf{X}(t | x)^2 \rangle\} - \{\langle \mathbf{X}(t | x) \rangle\}^2 \right)\\
	V(t | x) &=& \frac{d}{dt} \{ \langle \mathbf{X}(t | x)\rangle\} \rangle\}^2,	\label{eq_2.1.4b}
\end{eqnarray}
where $\mathbf{X}(t | x)$ denotes trajectories that originate at the \emph{radial} position $x$ at time $t = 0$ but with other space  $\mathbf{y}\equiv \left(\theta,\varphi\right)$ or velocity $\left(v_\parallel,\mu\right)$ coordinates unrestricted. Thus, $D(t | x)$ provides a local estimate of the diffusion coefficient at $x$. 


The formulas for transport coefficients (Eqs. \ref{eq_2.1.4a}-\ref{eq_2.1.4b}) involve two types of averages denoted by distinct brackets. The $\langle \cdot \rangle$ average is performed over an ensemble of random field realization and, as it will be proven in Section \ref{Section_3.5}, it is valid and motivated by the ergodicity of the turbulent transport. The  $\{ \cdot \}$ average is, in reality, an integral over the reminder of the phase-space coordinates without the radial $x$. Thus $\{ \cdot \}\equiv \int~J f(\mathbf{X},v_\parallel,\mu)dv_\parallel d\mu d\mathbf{y}$ is kinetic in nature and independent of turbulence. Note that $J=B_\parallel^\star$ is the Jacobian of the gyrocenter transformation and $\mathbf{y}$ is associated with the so-called flux-surface-average.


\subsection{The numerical code T3ST}
\label{Section_2.3}

The newly developed code T3ST \cite{palade2025t3st}, an acronym for \emph{Turbulent Transport in Tokamaks via Stochastic Trajectories}, is a numerical framework designed to compute charged particle trajectories in axisymmetric tokamak environments while statistically accounting for turbulent electrostatic fields. It directly implements the theoretical framework of particle transport outlined in the previous section, making it ideally suited for this study. Without delving into excessive detail, we highlight below the key features of T3ST that are relevant for the present analysis.

T3ST simulates the motion of an ensemble of $N_p$ test particles, each moving independently in its own realization of a turbulent electrostatic field. The particles’ initial phase-space distribution is customizable, although in this study, it is usually assumed a Maxwell-Boltzmann distribution in the velocity space, with particles uniformly distributed along the $(\theta, \varphi)$ directions on a magnetic flux tube at fixed radius $r = r_0$.

In each field realization, the synthetic turbulent potential is build with the aid of $N_c$ pairs of random wavenumbers and frequencies, $\{\mathbf{k}_i, \omega_i\}_{i,\in 1,N_c}$, sampled from a probability distribution function (PDF) defined by the normalized turbulence spectrum $S(\mathbf{k}, \omega)$. The chaotic nature of fluctuations is also controlled by random phases $\alpha_i \in [0, 2\pi)$ that are assigned to each mode. The resulting potential (evaluated at the gyrocenter) is computed numerically as:
\begin{eqnarray}
	\label{eq_2.2.1}
	\phi_1^{gc}(\mathbf{X}, t) = \sqrt{\frac{2}{N_c}} \sum_{i=1}^{N_c} J_0(k_i^\perp \rho_L) \sin\left(\mathbf{k}_i \cdot \mathbf{X} - \omega_i t + \alpha_i\right),
\end{eqnarray}
where the factor $\sqrt{2/N_c}$ ensures normalization, $J_0(k_i^\perp \rho_L)$ accounts for finite Larmor radius (FLR) effects due to gyroaveraging, and $k_i^\perp$ is the perpendicular component of $\mathbf{k}_i$ with respect to the local magnetic field. For further technical details, see \cite{Palade2021,palade2025t3st}.

To remain consistent with gyrokinetic conventions, field-aligned coordinates $(x, y, z)$ are used in the generation of turbulent fields \cite{Beer_fieldaligned}, defined as:
\begin{equation}
	\label{eq_2.2.2}
	x = C_x \rho(\psi) \approx r, \qquad y = C_y (\varphi - \bar{q} \chi), \qquad z = C_z \chi,
\end{equation}
where $C_x = a$ is the tokamak’s minor radius, $C_y = r_0 / \bar{q}(r_0)$ is a normalization based on a reference radius $r_0$, $C_z = 1$, and $\rho(\psi) \equiv \rho_t = \sqrt{\Phi_t(\psi)/\Phi_t(\psi_{edge})}$ is the normalized effective radius, approximated by $\rho_t \approx r/a$ in circular geometries. Here, $\Phi_t(\psi)$ is the toroidal magnetic flux $\Phi_t(\psi) = \int_{\psi_{axis}}^\psi \mathbf{B}\cdot\mathbf{e}_\varphi d S(\psi^\prime)$
measuring the flux enclosed between the axis and the surface labeled by $\psi$.

The turbulence spectrum $S(\mathbf{k}, \omega)$ used in this study captures key features of ITG-driven turbulence. It is constructed from analytical forms derived from saturation arguments and growth-rate-based heuristics \cite{10.1063/1.4954905,Dudding_2022}:
\begin{eqnarray}
	\label{eq_2.2.3}
	S(\mathbf{k}, \omega) &=& A_\phi^2 \frac{\tau_c \lambda_x \lambda_y \lambda_z}{(2\pi)^{5/2}} \frac{e^{-\frac{k_x^2 \lambda_x^2 + k_z^2 \lambda_z^2}{2}}}{1 + \tau_c^2 \omega^2} \frac{k_y}{k_0} \times\\
	&\times& \left( e^{-\frac{(k_y - k_0)^2 \lambda_y^2}{2}} - e^{-\frac{(k_y + k_0)^2 \lambda_y^2}{2}} \right),\nonumber
\end{eqnarray}
where $\lambda_x$, $\lambda_y$, and $\lambda_z$ are the spatial correlation lengths along the field-aligned directions $(x, y, z)$, $k_0$ sets the characteristic scale of the most unstable mode (influenced jointly with $\lambda_y$), $\tau_c$ is the correlation time, representing the departure of actual mode frequencies from linear predictions due to nonlinear interactions, and $A_\phi$ characterizes the turbulence amplitude. 

\subsection{Setup of numerical simulations}
\label{Section_2.4}

While T3ST solves equations of motion that require many physical parameters, we restrict here to the famous scenario called "Cyclone Base Case" (CBC) that corresponds to a typical DIII-D discharge \cite{A.M.Dimits_2000,Falchetto_2008}. The values of the relevant parameters are: $T_i=T_e=0.5keV, n_0=10^{19}m^{-3}, B_0 = 1.9T, R_0=1.71m, R_0/L_{T_i} =6.9, R_0/L_n=2.2, a= 0.625m, c_1 = 0.85, c_2 = 0, c_3 = 2.2, r_0 = a/2$. Note that, in the case of $H$ ions, $\rho_i/a=\rho^\star \approx 1/519$ and the values for the safety factor $q(r)=c_1+c_3(r/a)^2$ and magnetic shear $\hat{s}=d\ln\bar{q}/d\ln r$ are $\bar{q}(r_0) = 1.4, \hat{s} = 0.78$. Gyrokinetic simulations \cite{10.1063/1.1647136} of this scenario have shown that the ITG turbulence develops from the dominant instability and has approximately constant phase velocity $v_{ph}\approx V_\star \approx v_{th}\rho_i/L_n $, i.e. $\omega_{\mathbf{k}} \approx v_{ph}k_\theta$. Regarding the growing rates $\gamma$, they encompass the interval $[0,0.7]\rho_i^{-1}$, with a maximum at $k_\theta\rho_i\approx 0.3$. The turbulent amplitude at mid-radius in the saturated regime is $\Phi = eA_\phi/T_i\approx 1.1\%$ with a radial correlation length of $\lambda_r \approx 7\rho_i$ and a peaked spectrum in the poloidal wavenumber at $k_\theta\rho_i\approx 0.15$. Finally, the electrostatic potential is time-correlated \cite{10.1063/1.1647136} $\langle\phi_1(\mathbf{x},t)\phi_1(\mathbf{x},0)\rangle \propto \exp\left(-t/\tau_c\right)$ with $\tau_c\approx 1/\gamma\approx 3/\omega\approx 10\rho_i/v_{ph}$. For T3ST, these parameters translate into scaled values as $A_i = 1, \Phi =0.011,\lambda_x=7,\lambda_y=5,k_0=0.05,\tau_c = 4$ while $\lambda_z\to \infty$ is chosen (no ballooning or parallel fluctuations).

The present simulations are performed for ${}^{1}_{1}\mathrm{H}$ ions of the bulk plasma, thus, Maxwell-Botlzmann distributed with the temperature $T_i$. Unless otherwise specified, most simulations use $N_p=5\times 10^{5}$ test-particles, each field realization being constructed with $N_c = 10^2$ Fourier modes \eqref{eq_2.2.1}. The dynamics is followed over $t_{max}=60 R_0/v_{th}$ with a time-step of $\Delta t = t_{max}/N_t, N_t=1500$.

For brevity, in the Result section \ref{Section_3}, whenever transport coefficients are discussed they are evaluated at $x\equiv r=r_0$, and are denoted as $V(t)\equiv V(t|r_0), D(t)\equiv D(t|r_0)$.

\section{Results}
\label{Section_3}

\subsection{The dynamical scenario}
\label{Section_3.1}

Before answering questions about the Lagrangian features of turbulent transport it is important to have a qualitative view on the nature of particle dynamics driven by ITG turbulence.

\begin{figure}
	\centering
	\subfloat[Quiescent case: particles remain confined to narrow orbits. \label{fig_1_a}]{
		\includegraphics[width=0.9\linewidth]{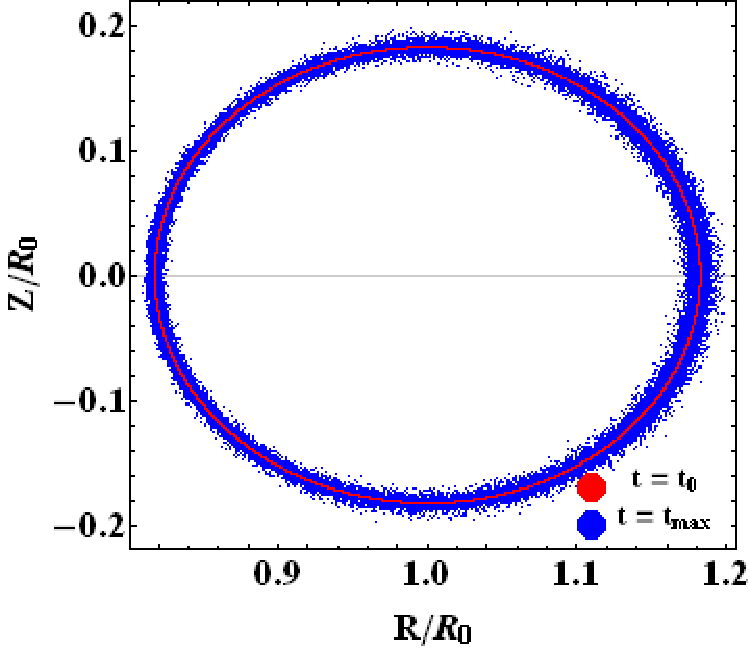}
	}\\
	\subfloat[Turbulent case: turbulence drives strong radial spreading, erasing FLR signatures.\label{fig_1_b}]{
		\includegraphics[width=0.9\linewidth]{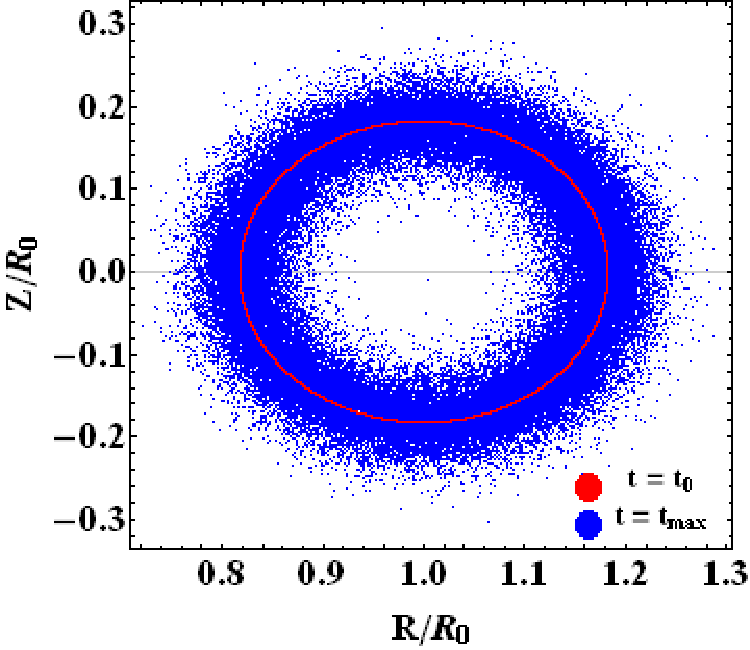}
	}
	\caption{Evolution of test-particle positions in the $(R, Z)$ poloidal plane. Red markers show initial positions ($t = 0$), and blue markers show final positions ($t = t_{max}$).}
\end{figure}

The numerical simulations of T3ST assume, as initial distribution of particles, Maxwell-Boltzmann statistics in the energy-pitch-angle space and uniform distribution over a thin flux tube defined by $r = r_0 = a/2$ in the physical space (a circle in poloidal projection). This corresponds to a \emph{local Maxwellian} \cite{10.1063/1.3519513}, which is not an equilibrium distribution—unlike the \emph{canonical Maxwellian} \cite{10.1063/1.3519513,10.1063/1.3153328,10.1063/1.2193947}. 

Consequently, even if particles are allowed to move solely under the influence of magnetic drifts (with no turbulence) the distribution function will evolve manifesting finite Larmor radius (FLR) effects. On the other hand, the quiescent particle trajectories are periodic orbits (banana or passing) which means they are effectively confined. One expects the system to reach, eventually, a steady-state with no radial transport.

This scenario should be broken once turbulence is introduced since the low-$k$ drift-type ITG imparts—mainly via the $\mathbf{v}_{E \times B}$ drift— correlated and continuous, but essentially random, kicks to particles. This should result in a non-equilibrium state with levels of transport that, hopefully, saturate asymptotically to finite values.

\begin{figure}
	\centering
	\subfloat[Radial running diffusion $D(t)$.\label{fig_2_a}]{
		\includegraphics[width=0.9\linewidth]{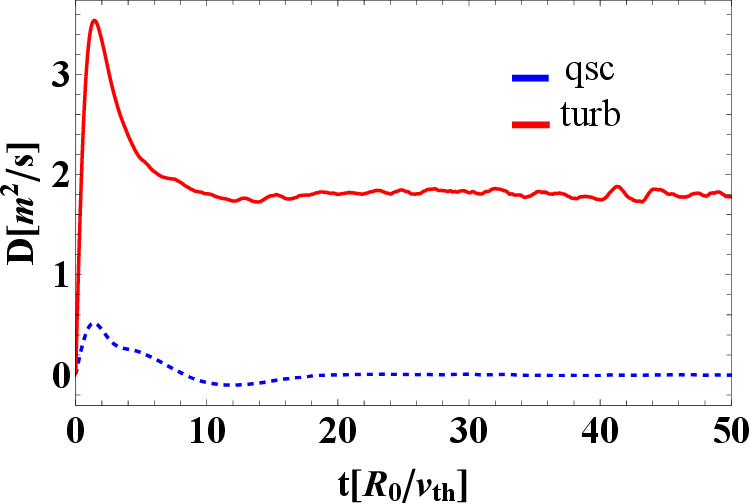}
	}\\
	\subfloat[Radial running velocity $V(t)$.\label{fig_2_b}]{
		\includegraphics[width=0.9\linewidth]{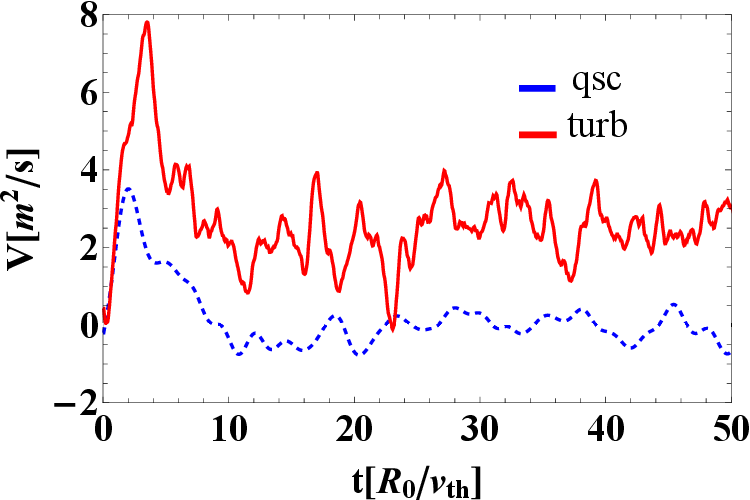}
	}
	\caption{Time evolution of radial transport coefficients. In both panels, blue lines represent quiescent dynamics (no turbulence), and red lines represent turbulent dynamics. Diffusion saturates under turbulence, while it vanishes in the quiescent case.}
\end{figure}

Numerical T3ST simulations are performed with or without turbulence. Figures \ref{fig_1_a} and \ref{fig_1_b} show, in the poloidal plane, the initial (red) and final ($t = t_{\text{max}}$, blue) particle distributions for the purely quiescent (a) and the turbulent cases (b). In the quiescent case (Fig.~\ref{fig_1_a}), FLR effects are evident: gyrocenters do not remain on the initial flux surface but spread along their orbits, producing a finite radial width—slightly broader on the low-field side. When turbulence is present (Fig.~\ref{fig_1_b}), particles undergo significant radial transport, effectively erasing the quiescent signature. 

\begin{figure}
	\centering
	\subfloat[Quiescent case.\label{fig_3_a}]{
		\includegraphics[width=0.9\linewidth]{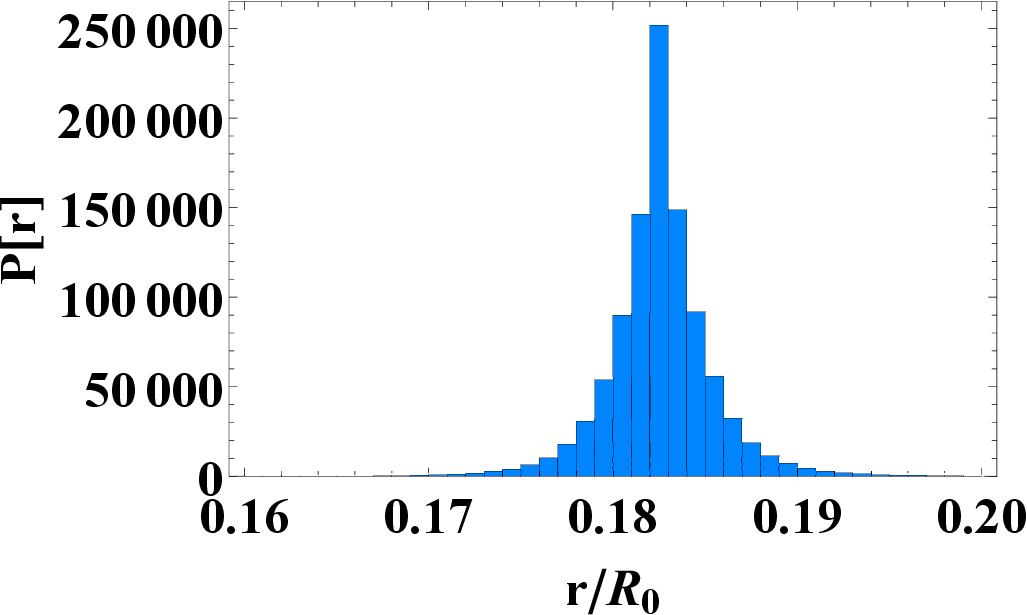}
	}\\
	\subfloat[Turbulent case.\label{fig_3_b}]{
		\includegraphics[width=0.9\linewidth]{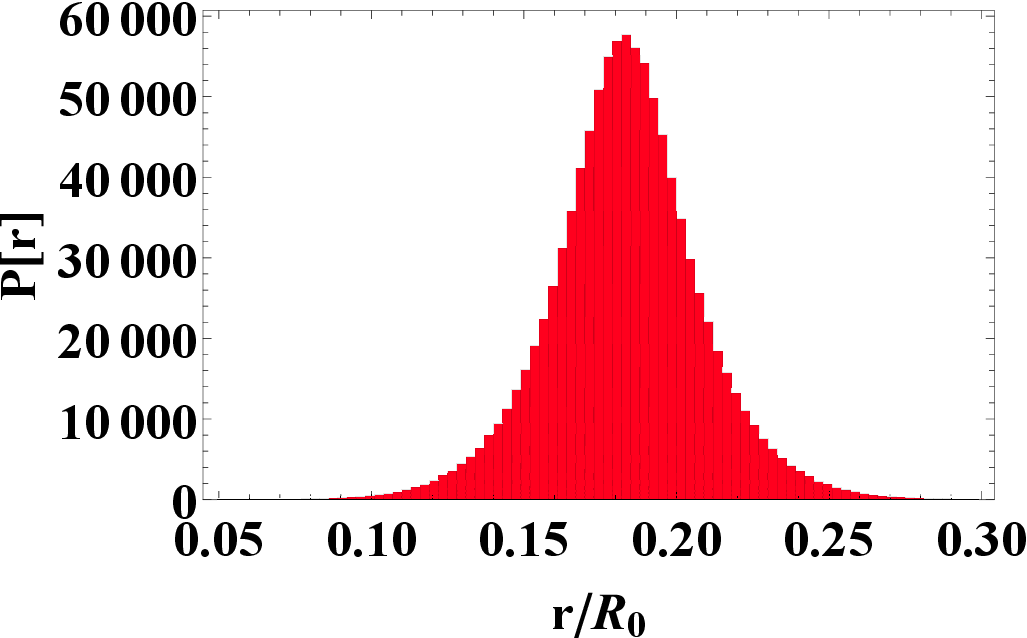}
	}
	\caption{Asymptotic ($t =t_{max}$) distributions of radial particle positions in the absence (Fig.~\ref{fig_3_a}, blue histogram) or the presence (Fig.~\ref{fig_3_b}, red histogram) of turbulence.}
\end{figure}

The expectation that quiescent motion leads asymptotically to confined equilibrium states with vanishing transport, while turbulence drives the system to non-equilibrium states with finite saturated transport is confirmed in Figs. \ref{fig_2_a} and \ref{fig_2_b}. These plots show the time-dependent transport coefficients: diffusion $D(t)$ (a) and convection $V(t)$ (b). After a short transient period ($t \sim 20 R_0 / v_{th}$), quiescent transport vanishes, while turbulent transport saturates to finite values. Notably, the convective term $V(t)$ is more susceptible to numerical fluctuations, despite both quantities being extracted from the same simulations.

\begin{figure}
	\centering
	\subfloat[Quiescent case: distribution retains long tails and symmetry.\label{fig_13_a}]{
		\includegraphics[width=0.9\linewidth]{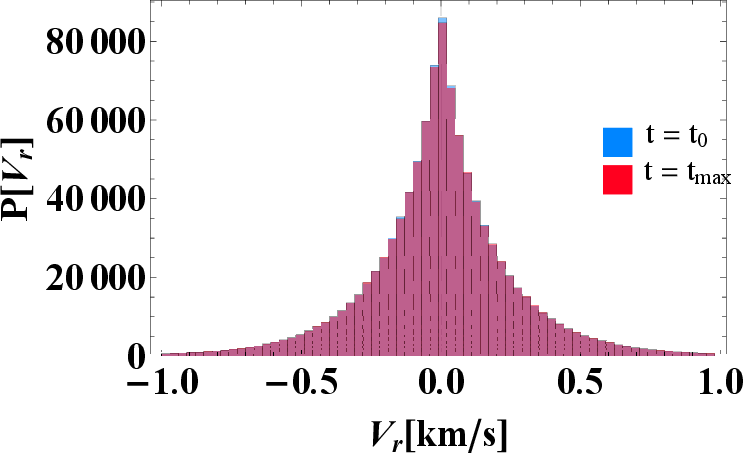}
	}\\
	\subfloat[Turbulent case: both initial and final distributions are nearly Gaussian and centered, consistent with diffusive behavior.\label{fig_13_b}]{
		\includegraphics[width=0.9\linewidth]{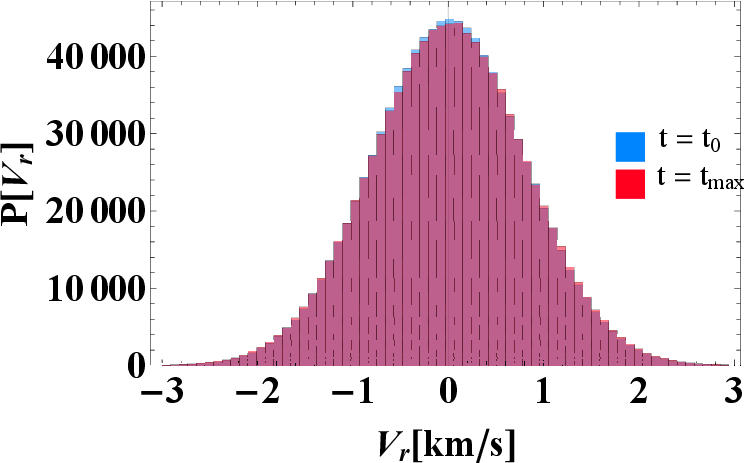}
	}
	\caption{Radial Lagrangian velocity distributions at initial (blue) and final (red) simulation times. }
\end{figure}

The diffusive character of turbulent transport is further evidenced by the radial particle distributions in Fig.~\ref{fig_3_b}, which approximates a Gaussian profile. In contrast, the quiescent case (Fig.~\ref{fig_3_a}) shows a narrower, symmetric distribution. These profiles can be understood by analyzing also the distributions of Lagrangian radial velocities which can be seen in Figs.~\ref{fig_13_a} and \ref{fig_13_b} at initial ($t = t_0$, blue) and final ($t = t_{\text{max}}$, red) times for the quiescent and turbulent cases, respectively. The nature of the profiles for velocities matches the profiles of radial tracers with or without turbulence. The reason for Gaussianity can be understood taking into account that turbulent drifts are, essentially, a sum of many random contributions (see Eq. \ref{eq_2.2.1} and the Central Limit Theorem). Conversely, the long tails in the quiescent case are attributed to the relatively simple dependence of magnetic drifts on a limited number of variables. A simple estimate shows that the radial magnetic drift velocity is 
\begin{eqnarray}
	\label{eq_3.1.1}
	V_r^{\text{neo}} \approx -\frac{m(v_\parallel^2 + v_\perp^2 / 2)}{q B^3} (\nabla B \times \mathbf{B}) \cdot \mathbf{e}_r\approx\\ \approx -\frac{(1 + \lambda^2)E}{q B_0 R_0} \sin\theta = -\frac{v_{th}\rho_i}{R_0}(1 + \lambda^2)\tilde{E} \sin\theta.\nonumber
\end{eqnarray}
Given that initially $\theta \in [0, 2\pi)$, $\lambda \in [-1, 1]$, and $P(\tilde{E}) \sim \sqrt{\tilde{E}} \exp(-\tilde{E})$, the resulting distribution can be approximated numerically as $P[V_r^{\text{neo}}, t=0]\approx \exp\left(-0.9 |V_r|R_0/(v_{th}\rho_i)\right)$. This is in good accordance with the numerical data from Fig. \ref{fig_13_a}.

What is remarkable is the fact that the velocity distributions seem to be extremely robust across dynamics with initial and asymptotic distributions matching almost perfectly (the overlapping of red and blue histograms results in a magenta hue, illustrating their near identity). This is a necessary condition for Lagrangian stationarity.

\subsection{The nature of the radial pinch}
\label{Section_3.2}

Figure~\eqref{fig_2_b} shows the effective running velocity coefficient $V(t)$, defined as the time derivative of the average radial position of the particles. Since this quantity is not constant over time—but instead exhibits a transient growth phase before reaching a stationary value—this implies that Lagrangian stationarity is invalid. But ain't this in direct contradiction with the apparent identical distributions of initial and final radial velocities shown in Fig. \ref{fig_13_b}? The answer is that a small displacement of $V(t\to t_{max})\approx 2m/s$ between distribution profiles in Fig. \ref{fig_13_b} is bellow the resolution of the figure, given that the MSD of velocities is $\sim 1km/s$, that is, three orders of magnitude higher. Note that this also explains the noisy character of $V(t)$.

\begin{figure}
	\centering
	\includegraphics[width=0.9\linewidth]{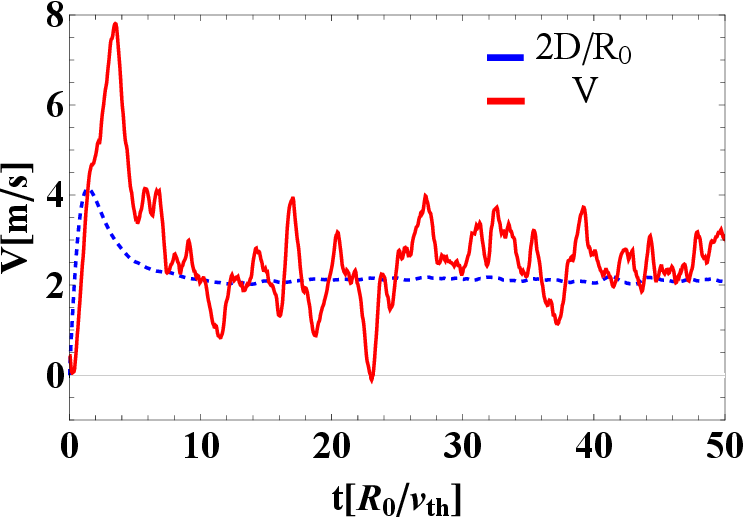}
	\caption{Effective velocity $V(t)$ (red, large fluctuations) compared to the normalized diffusion coefficient $2D(t)/R$ (blue, smoother behavior).}
	\label{fig_5}
\end{figure}

At early times, the radial pinch arises from both quiescent and turbulent effects, given that both cases exhibit a similar growing phase. The asymptotic value, on the other hand, is driven solely by turbulence, since the quiescent case leads to $V(t)\to 0 $ as $t\to t_{max}$. But what is the mechanisms behind the existence of this effective pinch? Currently, there are several pinch mechanisms identified in literature that rely on the existence of thermal, rotational \cite{Camenen_rotation}, magnetic field inhomogeneities \cite{Isichenko_invariant_pinch,Naulin_TEP,vlad_ratchet} or polarization drifts \cite{Palade_2023}. Since the present work does not include temperature gradients, toroidal rotation or polarization drift effects, all these are excluded. It remains possible that the pinch is a Turbulence Equipartition Pinch (TEP) \cite{Isichenko_invariant_pinch,Naulin_TEP,vlad_ratchet} which relies on the inhomogeneity of the magnetic field and on the compressibility of the $\mathbf{v}_{E\times B}$ drift. 

The nature of this effect can be emphasized with a simple perturbativ calculus, using the linearization $B(\mathbf{x})^{-1}\approx B(\mathbf{0})^{-1} - B(\mathbf{0})^{-2} \mathbf{x}\nabla B(\mathbf{0})$: 
\begin{align*}
	\mathbf{v}_{E\times B}(\mathbf{X}(t),t) = \frac{\mathbf{b} \times \nabla \phi(\mathbf{X}(t),t)}{B} \approx \\\approx \frac{\mathbf{b} \times \nabla \phi(\mathbf{X}(t),t)}{B(\mathbf{0})} \left(1 - \mathbf{X}(t) \cdot \nabla \ln B(\mathbf{0}) \right).
\end{align*}

Taking the ensemble average and considering a homogeneous distribution of positions $\mathbf{X}(t)$  that are driven mainly by the $E\times B$ drift, it yields:
\begin{align*}
	V(t)&=\left\langle \frac{\mathbf{b} \times \nabla \phi(\mathbf{X}(t),t)}{B(\mathbf{0})} \int_0^t d\tau \frac{\mathbf{b} \times \nabla \phi(\mathbf{X}(\tau),\tau)}{B(\mathbf{0})} \cdot \nabla \ln B(\mathbf{0}) \right\rangle \\
	& = \int_0^t d\tau \left\langle \frac{\mathbf{b} \times \nabla \phi(\mathbf{X}(t),t)}{B(\mathbf{0})} \frac{\mathbf{b} \times \nabla \phi(\mathbf{X}(\tau),\tau)}{B(\mathbf{0})} \right\rangle \cdot \nabla \ln B(\mathbf{0}) \\
	&\approx \int_0^t d\tau \left\langle V_r(t)V_r(\tau)\right\rangle \cdot \partial_r \ln B(\mathbf{0}) \propto \frac{2D(t)}{R_0}.
\end{align*}

In the final step we have identified the expression of diffusion as time-integral of the velocity auto-correlation. It turns out that the numerical results are very much in line with this approximate dependency (see Fig. \ref{fig_5}) thus, the TEP is confirmed.

\subsection{Lagrangian stationarity}
\label{Section_3.3}

Previously, results shown in Figs. \ref{fig_13_a}-\ref{fig_13_b} suggested that the distributions $P[V_r]$ of radial Lagrangian velocities of particles are almost identical between the starting point of the simulation $t=0$ and the final time $t=t_{max}=60R_0/v_{th}$. A closer inspection into Fig. \ref{fig_5} has revealed that this is not entirely true: particles do experience an average Lagrangian velocity $V(t)$ which is of TEP nature and results from the inhomogeneity of the magnetic field. It is not visible in the plot of distributions due to scale disparity: $V(t)=\langle V_r(t)\rangle \sim 1m/s$, while $\sqrt{\langle V_r^2(t)\rangle }\sim 1km/s$. Thus, we conclude that stationarity is broken for the average of velocities but is approximately valid in the asymptotic region.

\begin{figure}
	\centering
	\subfloat[Quiescent case.\label{fig_6_a}]{
		\includegraphics[width=0.9\linewidth]{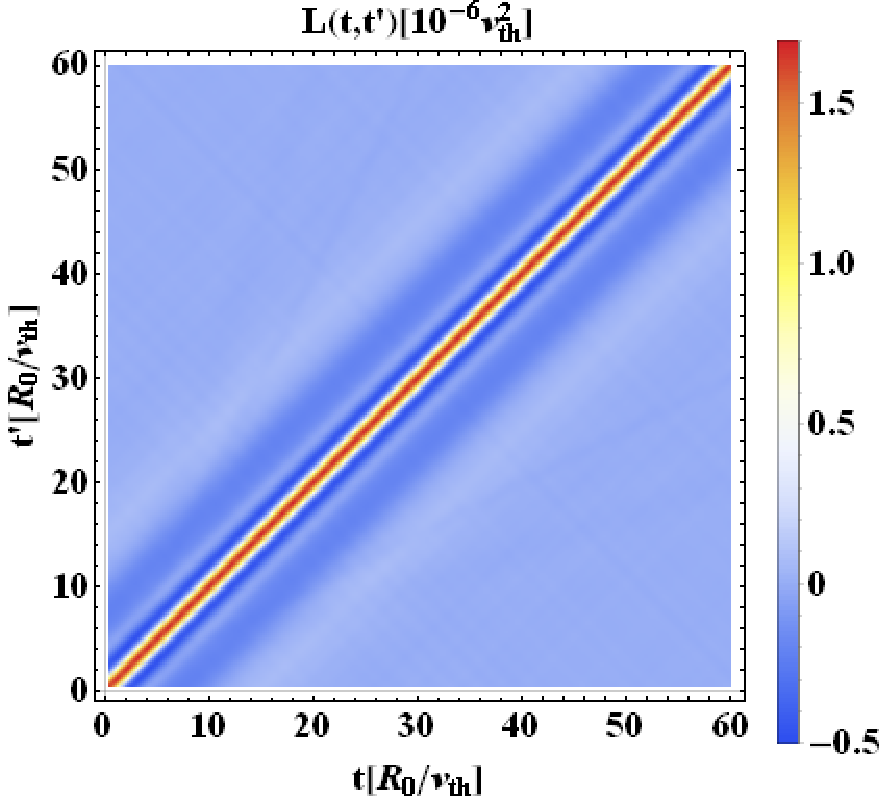}
	}\\
	\subfloat[Turbulent case.\label{fig_6_b}]{
		\includegraphics[width=0.9\linewidth]{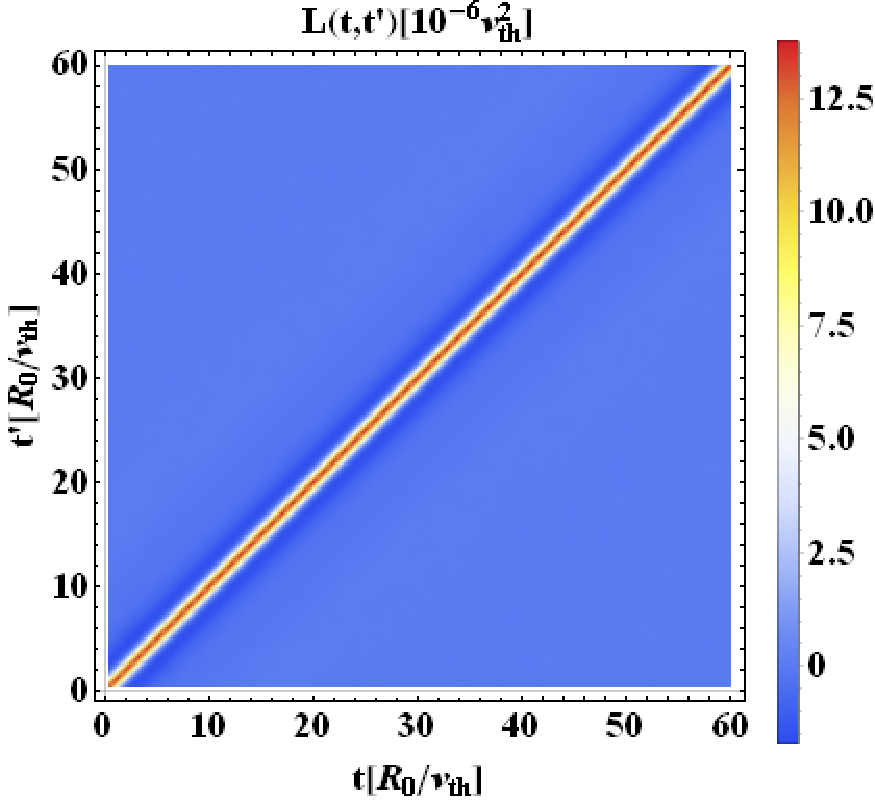}
	}
	\caption{Lagrangian auto-correlation $L(t,t^\prime)$ of radial velocity fields in the quiescent (a) and the turbulent (b) cases.}
\end{figure}

We look further to the Lagrangian velocity auto-correlation along the radial direction, defined, in this work's notations, as:

\begin{eqnarray}
L(t,t^\prime) = \{\langle V_r(t)V_r(t^\prime)\rangle\}-\{\langle V_r(t)\rangle\}\{\langle V_r(t^\prime)\rangle\}	\label{eq_3.3.1}
\end{eqnarray}

\begin{figure}
	\centering
	\subfloat[Quiescent case.\label{fig_7_a}]{
		\includegraphics[width=0.9\linewidth]{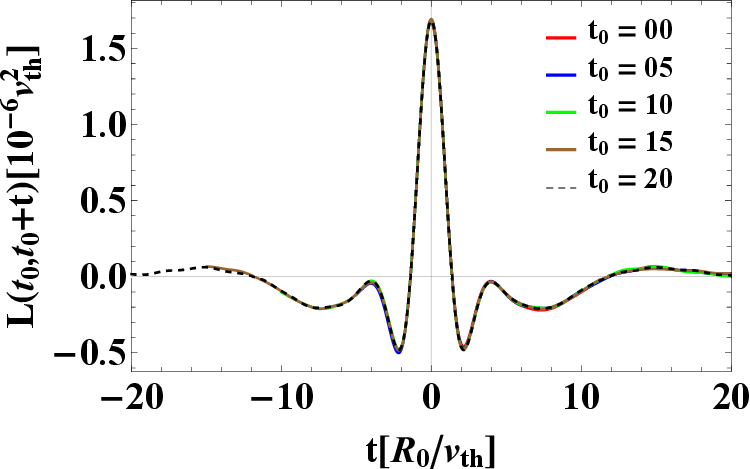}
	}\\
	\subfloat[Turbulent case.\label{fig_7_b}]{
		\includegraphics[width=0.9\linewidth]{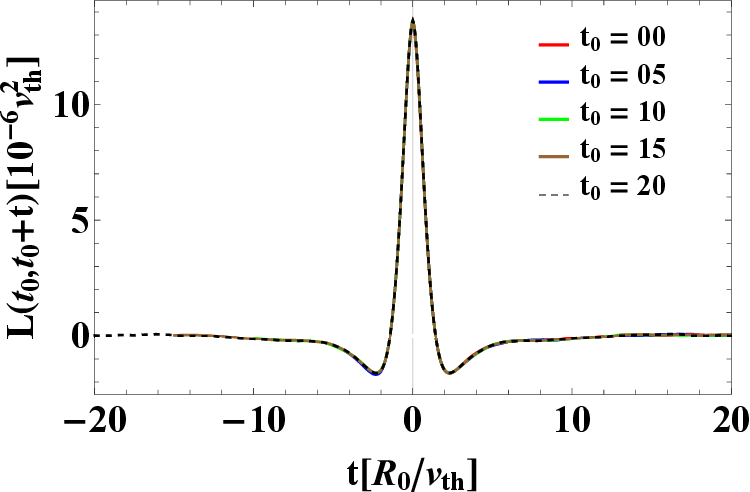}
	}
	\caption{Lagrangian auto-correlation $L(t_0,t_0+t)$ evaluated in the quiescent (a) and the turbulent (b) cases for $t_0 = 0,5,10,15,20 R_0/v_{th}$ (red, blue, green, brown, respectively black lines). The curves are essentially indistinguishable.}
\end{figure}

\begin{figure}
	\centering
	\subfloat[Quiescent case.\label{fig_8_a}]{
		\includegraphics[width=0.9\linewidth]{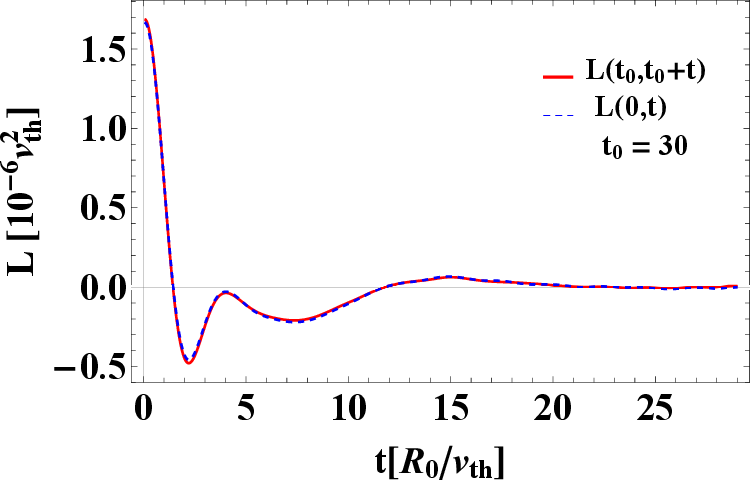}
	}\\
	\subfloat[Turbulent case.\label{fig_8_b}]{
		\includegraphics[width=0.9\linewidth]{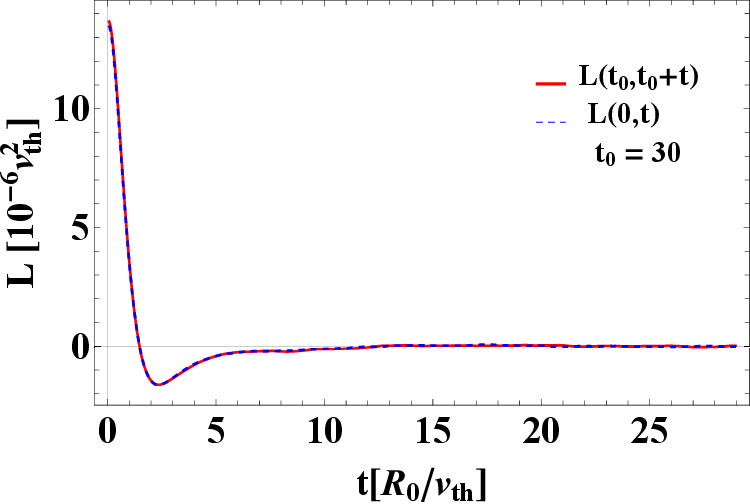}
	}
	\caption{Lagrangian auto-correlation $L(t_0,t_0+t)$ evaluated in the quiescent (a) and the turbulent (b) cases for $t_0 = 0,20 R_0/v_{th}$ (blue, red lines). The curves are hardly distinguishable.}
\end{figure}

If the dynamics would be truly stationary, this quantity should be time-invariant, i.e. $L(t,t^\prime)\equiv L(|t-t^\prime|,0)$. In Figs. \ref{fig_6_a}-\ref{fig_6_b} is plotted precisely $L$ for the quiescent (a) and the turbulent case (b). It appears that, at least qualitatively, the graphs are in both cases symmetrical, implying stationarity. The matter can be explored further by investigating slices of $L(t_0,t_0+t)$ in terms of the time-difference $t$. This is shown for many $t_0$ values in Figs. \ref{fig_7_a}-\ref{fig_7_b} (quiescent and turbulent case) and in Figs. \ref{fig_8_a}-\ref{fig_8_b} for only two values ($t_0 = 0$ - blue line and $t_0=30$ - red line). In all these figures the lines are virtually indistinguishable, thus, signalling almost perfect stationarity of the Lagrangian velocities across the super-ensemble.

\begin{figure}
	\centering
	\includegraphics[width=0.9\linewidth]{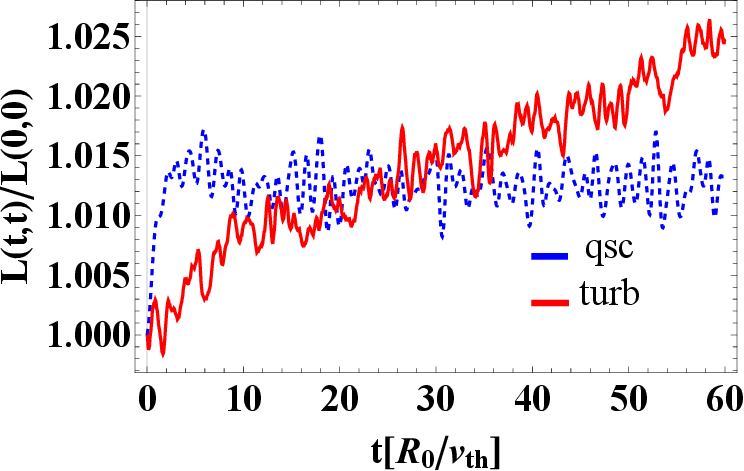}
	\caption{Time evolution of the second moment of the distribution of velocities scaled to its initial value $\{\langle V^2(t)\rangle\} - \{\langle V(t)\rangle\}^2 = L(t,t)$ for the quiescent (blue) and the turbulent (red) cases.}
	\label{fig_4}
\end{figure}

We now go back to the local in time dynamics and ask weather the second-order cumulant of Lagrangian velocities is time-invariant. So it happens that this quantity is identical to the diagonal part of the correlation function, i.e. $\{\langle V^2(t)\rangle\} - \{\langle V(t)\rangle\}^2 = L(t,t)$. The results are shown in Fig. \ref{fig_4} where small departures from stationarity, i.e. $\sim 1\%$ can be observed both for the quiescent (blue) and the turbulent (red) cases. While in the former case one observes only a transient growth followed by a saturation, the latter exhibits continuous linear growth. This again must be connected with the inhomogeneity of $B$ and it suggests that the transport might not even be perfectly saturated (or local, for that matter). 

Thus, given the results from this section, one can conclude that the Lagrangian stationarity is approximately present, with small deviations that are, in general, quantifiable to $\approx 1\%$. 

A more surprising aspect is that there is stationarity in the quiescent case. This can't be attributed to the approximate homogeneity of the turbulence and it is in apparent striking conflict with the fact that magnetic drifts \ref{eq_3.1.1} are starkly inhomogeneous. The only player in this picture that could drive Lagrangian stationarity are the particles, more precisely, their guiding-center coordinates. Indeed, the initial phase-space distribution function used by T3ST although it is one of non-equilibrium, it evolves conserving the phase-space volume. Given how wide it is, it must be the reason behind Lagrangian homogeneity and ergodicity.

\subsection{Statistics of field derivatives}
\label{Section_3.6}

The Lagrangian stationarity of the velocity statistics was proven to be approximately true, despite the inhomogeneous and compressible nature of the Eulerian drift field. A natural extension of this analysis is to investigate whether the \textit{Lagrangian statistics of the potential gradients}, which directly generate the turbulent $\mathbf{E} \times \mathbf{B}$ drift, also exhibit stationary behavior over time.

In the gyrokinetic approximation, the dominant turbulent contribution to particle motion arises from the electrostatic potential via the drift:
\begin{equation}\label{eq_3.4.1}
	\mathbf{v}_{E \times B} = \frac{\mathbf{b} \times \nabla \phi}{B},
\end{equation}
where $\phi$ is the fluctuating electrostatic potential evaluated at the gyrocenter. Therefore, the gradients $\nabla \phi$—and in particular, their Lagrangian statistics—are key drivers of transport.

We compute the first- and second-order Lagrangian statistics of the field derivatives over time, namely:
\begin{equation}\label{eq_3.4.2}
	\left\langle \partial_i \phi(t) \right\rangle, \quad \text{and} \quad \left\langle \left( \partial_i \phi(t) \right)^2 \right\rangle, \quad i \in \{x, y\},
\end{equation}
where the spatial derivatives are evaluated along test-particle trajectories and averaged over the super-ensemble. 

\begin{figure}
	\centering
	\subfloat[Derivative's average.\label{fig_14_b}]{
		\includegraphics[width=0.9\linewidth]{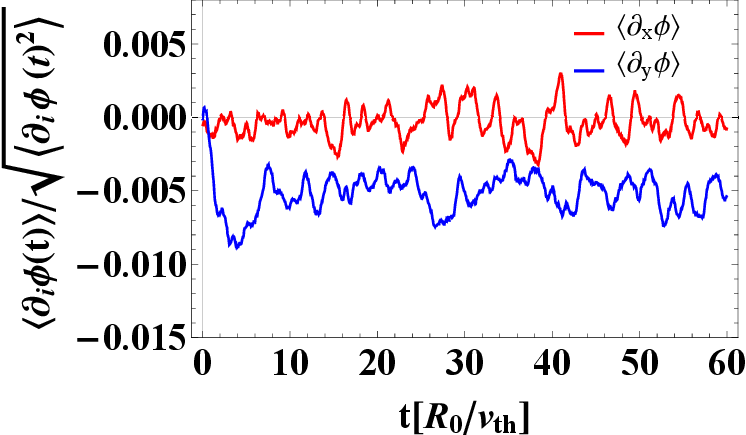}
	}\\
	\subfloat[Derivative's amplitude.\label{fig_14_a}]{
		\includegraphics[width=0.9\linewidth]{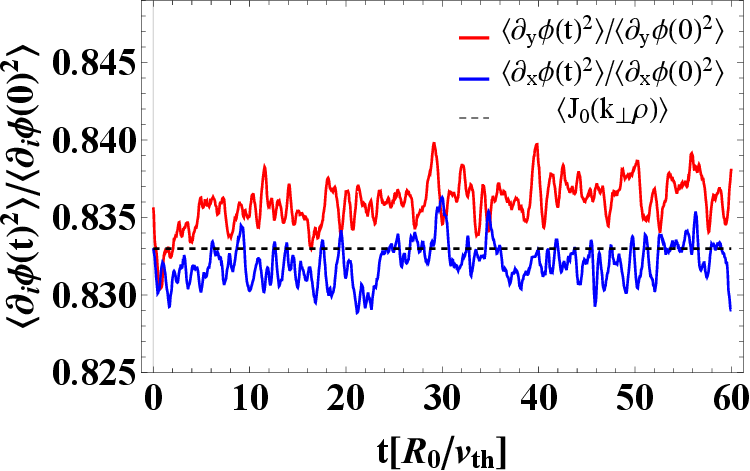}
	}\\
	\subfloat[Poloidal derivative distribution.\label{fig_14_c}]{
		\includegraphics[width=0.9\linewidth]{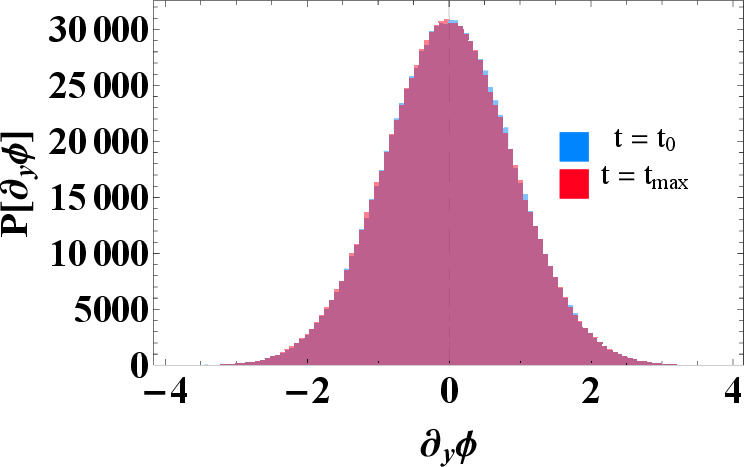}
	}
	\caption{Time evolution of the Lagrangian average of field derivatives $\{\langle\partial_x\phi(t)\rangle\},\{\langle\partial_y\phi(t)\rangle\}$ (red,blue) (\ref{fig_14_b}) and derivative's amplitudes $\{\langle\left(\partial_x\phi(t)\right)^2\rangle\},\{\langle\left(\partial_x\phi(t)\right)^2\rangle\}$ (red,blue) (\ref{fig_14_a}). In fig. \ref{fig_14_c} is shown the distribution of poloidal derivatives $\partial_y\phi(t)$ at the initial (blue, $t=t_0$) and the final (red, $t=t_{max}$) simulation times.} 
\end{figure}

Figure~\ref{fig_14_b} shows the time evolution of the average values $\left\langle \partial_x \phi \right\rangle$ and $\left\langle \partial_y \phi \right\rangle$, normalized by their respective standard deviations. Both quantities remain very close to zero throughout the simulation time, indicating that there is no net directional bias in the turbulent forcing fields along particle paths. This is consistent with the Eulerian property that the turbulent potential has zero mean and is symmetric in space.

Figure~\ref{fig_14_a} presents the normalized second moments $\left\langle \left( \partial_i \phi \right)^2 \right\rangle / \left\langle \left( \partial_i \phi \right)^2 \right\rangle_{t=0}$, which represent the average amplitude of the turbulent gradients as experienced along Lagrangian trajectories. These amplitudes remain stable over time, with only small fluctuations (below $2\%$) relative to their initial values. This indicates that the \textit{turbulence maintains its effective strength along the paths of particles}, and supports the earlier observation of approximate Lagrangian stationarity in drift velocities.

Additionally, Figure~\ref{fig_14_c} compares the probability distributions $P[\partial_y \phi]$ ($\partial_y \phi$ is the main contribution to radial transport) at the initial and final simulation times. The distributions are nearly indistinguishable and closely approximate Gaussian profiles, consistent with the assumption of normally distributed fluctuations in $\phi$. The invariance of these distributions over time provides further evidence for the ergodic and statistically stationary character of the turbulent forcing fields along gyrocenter trajectories.

Taken together, these results reinforce the conclusion that \textit{not only the Lagrangian velocities but also the turbulent driving gradients remain statistically stationary over time}, despite the fact that the Eulerian fields themselves are inhomogeneous and compressible. 

\subsection{Time-symmetry}
\label{Section_3.7}

Lagrangian stationarity, in the general case, does not require nor it implies time-reversibility/symmetry. The applicability of the Lagrangian method for the calculation of diffusion ($D_L(t)$, see Section \ref{Section_3.4}), on the other hand, requires symmetry since the Lagrangian velocity auto-correlation must obey $L(t,t^\prime)= L(t-t^\prime,0)=L(t^\prime-t,0)$. For this reason, in this section the time-symmetry of turbulent transport is investigated .

\begin{figure}
	\centering
	\subfloat[Running diffusion coefficient.\label{fig_16_a}]{
		\includegraphics[width=0.9\linewidth]{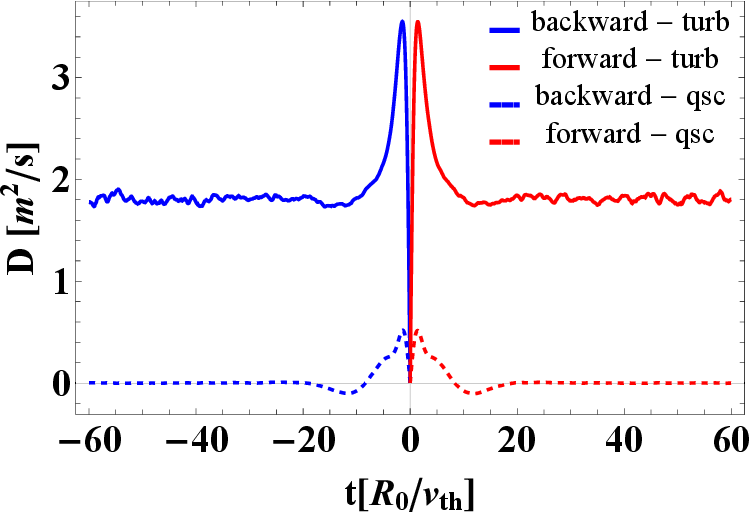}
	}\\
	\subfloat[Running velocity coefficient.\label{fig_16_b}]{
		\includegraphics[width=0.9\linewidth]{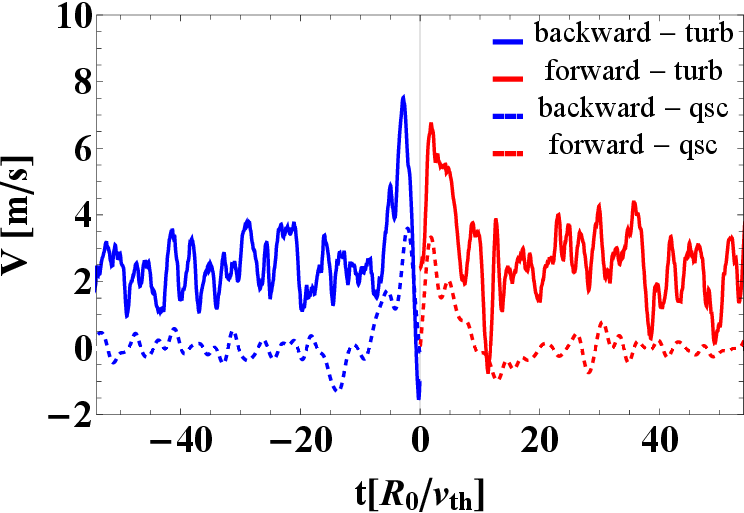}
	}
	\caption{Running diffusion (a) and velocity (b) coefficients for quiescent (dashed lines) and turbulent (solid lines) dynamics, computed forward (red) and backward (blue) in time.}
\end{figure}

\begin{figure}
	\centering
	\subfloat[Quiescent case.\label{fig_16_d}]{
		\includegraphics[width=0.9\linewidth]{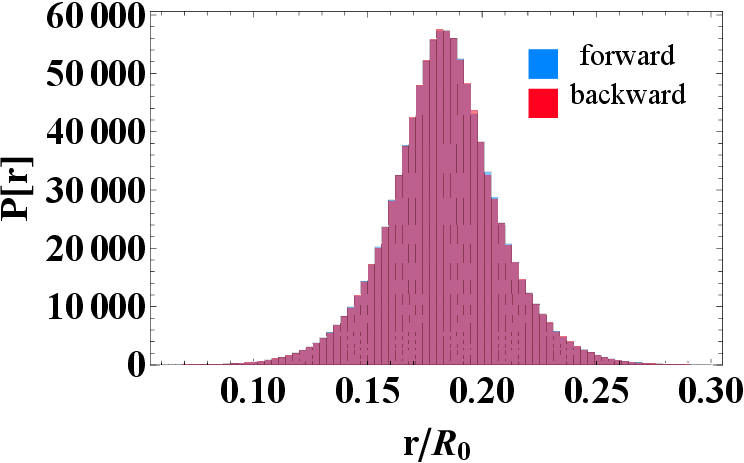}
	}\\
	\subfloat[Turbulent case.\label{fig_16_c}]{
		\includegraphics[width=0.9\linewidth]{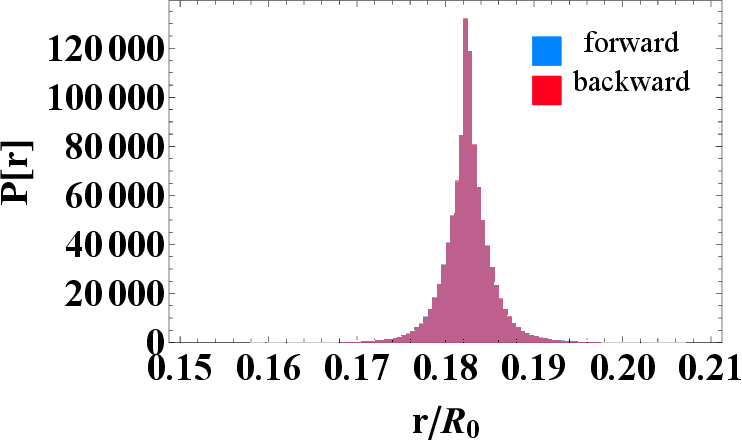}
	}
	\caption{Long-time ($t = t_{\text{max}}$) distribution of particle radial positions for quiescent (a) and turbulent (b) dynamics, computed forward (red) and backward (blue) in time.}
\end{figure}

\begin{figure}
	\centering
	\includegraphics[width=0.9\linewidth]{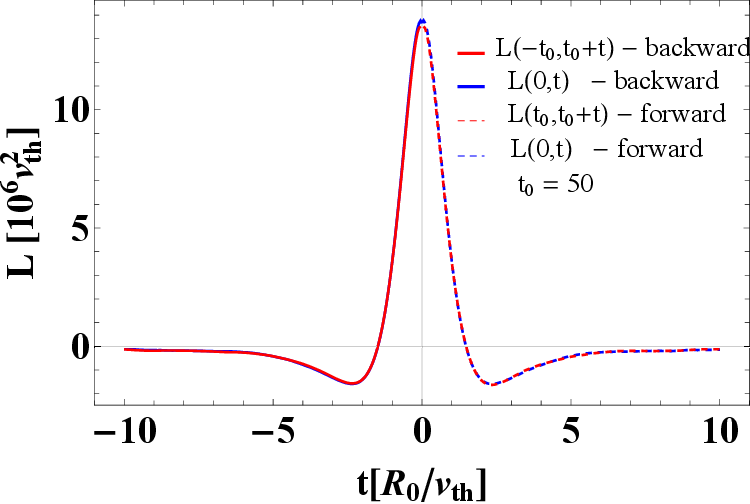}
	\caption{Lagrangian velocity autocorrelation $L(t_0, t_0 + t)$ for the turbulent case, evaluated at $t_0 = 0$ and $t_0 = 20 R_0 / v_{th}$ (blue and red lines), for forward (dashed) and backward (solid) dynamics.}
	\label{fig_16_e}
\end{figure}

The time direction in the numerical integration of particle trajectories is simply inverted. We then compare transport quantities: diffusion (Fig.~\ref{fig_16_a}) and velocity (Fig.~\ref{fig_16_b}) coefficients, radial particle distributions (Figs.~\ref{fig_16_d}--\ref{fig_16_c}), and the Lagrangian velocity autocorrelation (Fig.~\ref{fig_16_e}).

Figs.~\ref{fig_16_a}--\ref{fig_16_b} show that forward (red) and backward (blue) transport coefficients are nearly symmetric with respect to $t = 0$, for both quiescent and turbulent dynamics. This symmetry, apart from small numerical fluctuations, indicates that the transport is effectively time-reversible.

This conclusion is further supported by Figs.~\ref{fig_16_c}-\ref{fig_16_d}, that present the final distributions of radial particle positions. They are nearly identical in the forward and backward simulations, indicating an even stronger manifestation of time-symmetric dynamics. Finally, Fig.~\ref{fig_16_e} shows that the Lagrangian velocity autocorrelation $L(t_0, t_0 + t)$ also exhibits the same symmetry between forward (dashed) and backward (solid) time evolutions.

These findings are consistent with the structure of the equations of motion (Eqs.~\ref{eq_2.1.1a}-\ref{eq_2.1.1b}). The magnetic drifts are time-independent, while the turbulent fields introduce time dependence through their mode frequencies $\omega = \omega_\star (\mathbf{k})+\Delta\omega$. Part of this time dependence $\Delta \omega$ arises from nonlinear saturation processes, which contribute symmetrically in time and are governed by the decorrelation time $\tau_c$. These components do not break time-reversibility. The dispersive part $\omega_\star(\mathbf{k})$ imposes a preferential direction in space and time for the drift of turbulent waves. However, the plasma equilibrium and particle distributions are space-time symmetrical relative to this special direction of drift, thus, inverting time switches the ITG into a TEM-like instability but without affecting the radial transport.

\subsection{On the validity of the statistical approach}
\label{Section_3.5}

As detailed in the Introduction, the use of statistical ensembles to study turbulent transport can be motivated by the \emph{epistemic} argument that turbulence is chaotic and its configuration cannot be precisely known. The rigorous argument is rather \emph{ontic} and relies on ergodicity that arises from space-homogeneity, time-stationarity and incompressibility of the Eulerian velocity field $\mathbf{v}(\mathbf{x},t)$. Given that these properties are not perfectly met by the gyrocenter drifts in tokamaks, we ask here weather a statistical ensemble of turbulent potentials $\phi(\mathbf{x},t)$ is able to produce transport features similar to a single field realization.

\begin{figure}
	\centering
	\includegraphics[width=0.9\linewidth]{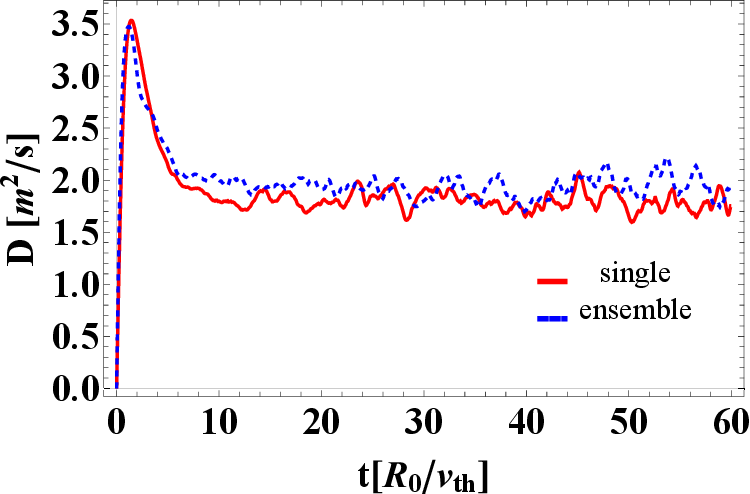}
	\caption{Running diffusion coefficient for the turbulent regime, evaluated for a single field 
	realization (red, solid line) and for a statistical ensemble of realizations (blue, dashed line).}
	\label{fig_12}
\end{figure}

Two numerical simulations are performed. In the first, $N_p$ particles evolve in a single turbulent field realization (shared by all particles); in the second, each of the $N_p$ particles evolves in its own independent realization of the turbulent field. The resulting diffusion coefficients for both cases are shown in Fig.~\ref{fig_12}. Minor differences $\sim 5\%$ can be observed across the entire time profile and stem from numerical fluctuations. It is interesting to note that the latter seem to have similar amplitudes in both cases. This suggests that the numerical noise is independent of the number of realizations. 

Thus, at least from the perspective of transport, modelling turbulence via an ensemble of random fields is equivalent to using a single realization of a chaotic field.

\subsection{The influence of initial distributions}
\label{Section_3.8}

Up to this point, all numerical results have indicated approximate Lagrangian stationarity of the turbulent dynamical processes, despite the Eulerian inhomogeneity of magnetic fields and $\mathbf{E}\times\mathbf{B}$ drifts. This behavior was attributed to the space-time ergodicity of particle trajectories, which in practice is supported by the broad initial distribution of particles in phase space which induce ergodic mixing. This hypothesis will be tested in this and the next section.

Here, we focus specifically on the impact of initial conditions on transport—particularly on the computed diffusion coefficients. Beyond its relevance to Lagrangian stationarity, this topic is important for a fundamental reason: in deriving Fick-like transport laws, $\Gamma = V n -D\nabla n$, whether from Onsager symmetry relations or Green-Kubo relations, it is generally assumed that the distribution function is either an equilibrium or a statistical average. 

However, in T3ST, particles are typically initialized uniformly over a flux-tube with a Maxwell-Boltzmann velocity distribution. This does not represent an equilibrium distribution, nor does it reflect the dynamical statistical average. A more consistent approach would be to initialize particles in either a known equilibrium state (such as a canonical Maxwellian) or a steady-state—like the one reached asymptotically under pure magnetic motion. Unfortunately, both alternatives would require additional and non-trivial numerical procedures.


To explore the sensitivity of transport to initial phase-space distributions, we carry out several comparative numerical simulations which differ from the typical simulation described in Section \ref{Section_2.4} trough one of the following:

\begin{enumerate}
	\item Initial pitch angles are fixed at $\lambda = 0.3$.
	\item Initial energies are fixed at $E = T_i$.
	\item Particles are placed at a single radial point on the low-field side (LFS).
	\item Turbulence is later, at $t = 35R_0/v_{th}$, after the particles reach a quasi-steady quiescent state.
	\item Initial pitch angles and energies are concurrently setted to $\lambda=0.3, E=T_i$. 
	\item Initial pitch angles are setted to $\lambda=0.3$ and particle placed at the low-field-side (LFS).
	\item Initial energies are setted to $E=T_i$ and particle placed at the low-field-side (LFS).
\end{enumerate}

\begin{figure}
	\centering
	\subfloat[Particles distributed initially in the standard configuration (black), placed at the low field side (blue), with a single energy (green), with a single pitch angle (brown) or with turbulence starting later ($t=35$ - red).\label{fig_15_e}]{
		\includegraphics[width=0.9\linewidth]{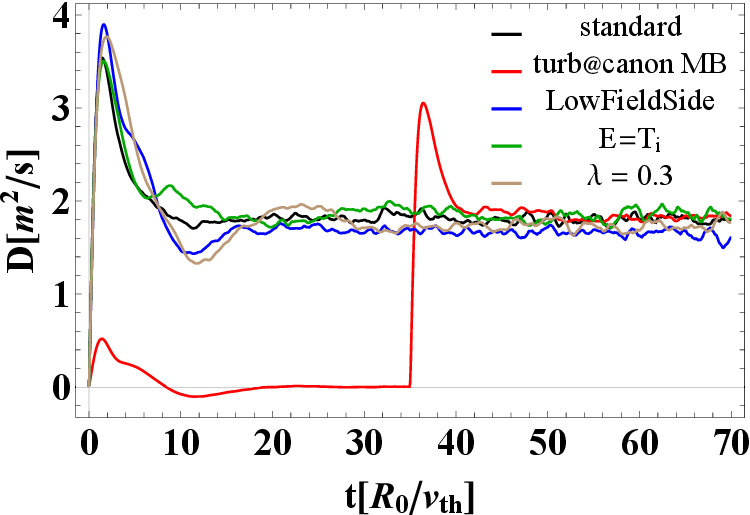}
	}\\
	\subfloat[Particles distributed initially in the standard configuration (black), placed at the low field side with a single energy (green) or with a single pitch angle (blue) and with a single energy and pitch angle (red).\label{fig_15_f}]{
		\includegraphics[width=0.9\linewidth]{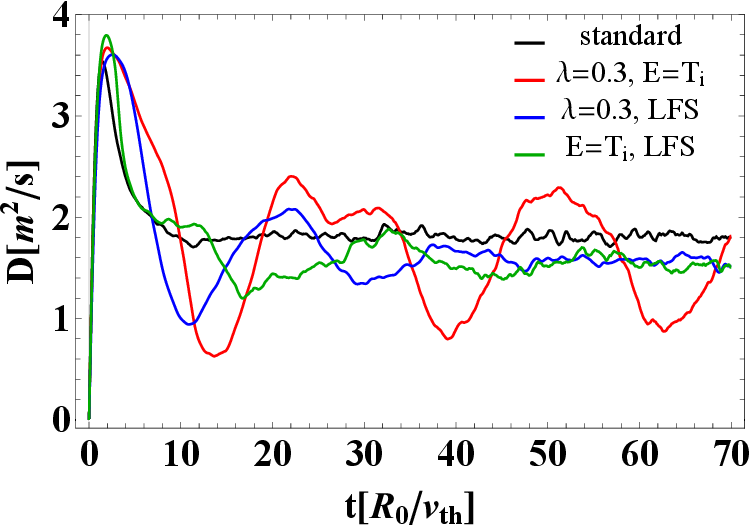}
	}
	\caption{Comparison of running radial diffusion coefficients obtained under different initial conditions.}
\end{figure}

The first four scenarios constitute a mild degradation (restriction) of the initial filling of the phase space. Their associated radial diffusion coefficients are shown in Fig.~\ref{fig_15_e}. Perhaps surprisingly, aside from short-time transients, the resulting asymptotic diffusion coefficients are largely insensitive to the choice of initial conditions. The only notable deviation occurs when particles are initialized exclusively at the low-field-side line, which leads to a modest ($\sim 5\%$) change in the long-time diffusion coefficient.

The most significant result shown in Fig.~\ref{fig_15_e} is that the diffusion coefficients are virtually identical regardless of whether turbulence is activated at $t=0$ (black) or later, at $t=35R_0/v_{th}$ (red). This indicates that initializing turbulence on a non-equilibrium particle distribution (the standard T3ST scenario) yields essentially the same asymptotic transport as starting from a near-equilibrium distribution. This is particularly valuable, as it justifies the use of the simpler and computationally cheaper red scenario.

We proceed further to the last three cases which are a consistent degradation of the initial phase-space occupied volume by the particles. The results are shown in Fig.~\ref{fig_15_f}. The time-profiles of diffusions and their asymptotic values are somehow more dispersed but they still surprisingly close ($\sim 20\%$). This tells us that even a modest initial filling of the phase-space leads to similar transport as much more extended distributions (the standard case).

\subsection{Two methods of computing diffusion}
\label{Section_3.4}

If the Lagrangian velocity auto-correlation function is indeed stationary—as approximately suggested in Section~\ref{Section_3.5}—i.e., $L(t, t^\prime) \equiv L(|t - t^\prime|,0)$, then the following definitions of the diffusion coefficient should be equivalent, $D_d(t) = D_L(t)$:
\begin{eqnarray}
	\label{eq_3.8.1}
	D_d(t) &=& \frac{1}{2} \frac{d}{dt}\left\langle \delta x^2(t) \right\rangle \\
	D_L(t) &=&  \int_0^t L(\tau) \, d\tau.
\end{eqnarray}

This equivalence is important for many theoretical and computational approaches. In particular, it underlies the Decorrelation Trajectory Method (DTM) \cite{PhysRevE.58.7359}, which has been widely applied in the study of turbulent transport in fluids \cite{PhysRevE.63.066304}, tokamaks \cite{PhysRevE.61.3023,Palade_2021_W,Vlad_2021}, and astrophysical plasmas \cite{vlad2018effects,negrea2017stochastic}. An alternative view on the matter can be drawn from the fact that $D_L(t)$ is, in reality, a Green-Kubo relation which stems from the fluctuation-dissipation theorem \cite{R_Kubo_1966}. The latter requires ergodicity and time-translational invariance \cite{vainstein2006mixing}.

\begin{figure}
	\centering
	\includegraphics[width=0.9\linewidth]{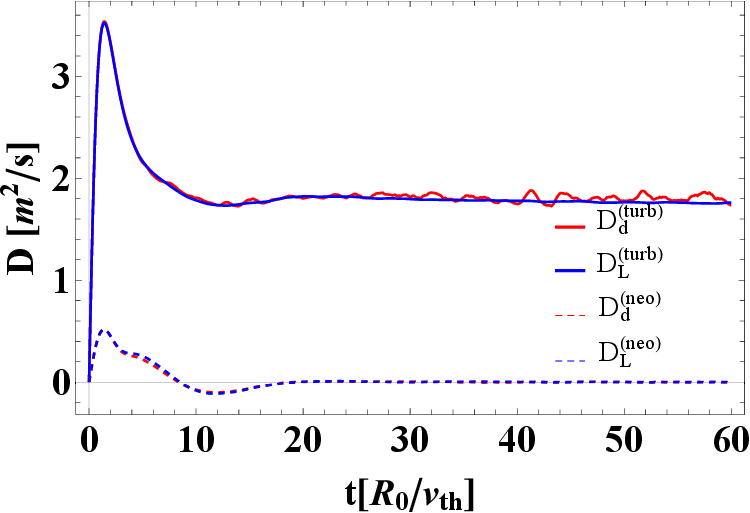}
	\caption{Running diffusion coefficients computed using the MSD-based method $D_d$ (red) and the Lagrangian correlation method $D_L$ (blue), for both the quiescent (dashed lines) and turbulent (solid lines) cases.}
	\label{fig_9}
\end{figure}

We assess the validity of the equality $D_d(t) = D_L(t)$ by comparing numerical data. They are shown in Fig.~\ref{fig_9}, where the MSD-derived diffusion coefficient $D_d^{(X)}(t)$ is plotted in red, and the correlation-based estimate $D_L^{(X)}(t)$ in blue for $X=$ neo (dashed lines) and $X=$ turb (filled lines). The agreement between the two methods is remarkably close, with only small $\sim 1\%$ discrepancies. This suggests, once again, good but not perfect stationarity of the Lagrangian correlation function $L(t, t')$. 

\begin{figure}
	\centering
	\subfloat[Quiescent case.\label{fig_10_a}]{
		\includegraphics[width=0.9\linewidth]{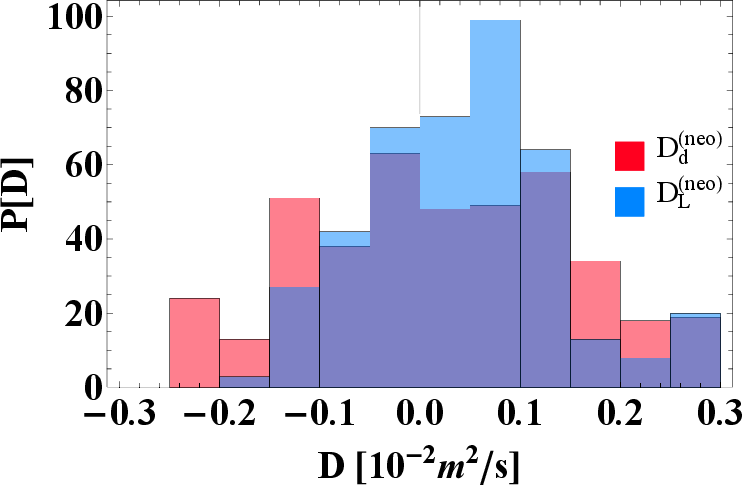}
	}\\
	\subfloat[Turbulent case.\label{fig_10_b}]{
		\includegraphics[width=0.9\linewidth]{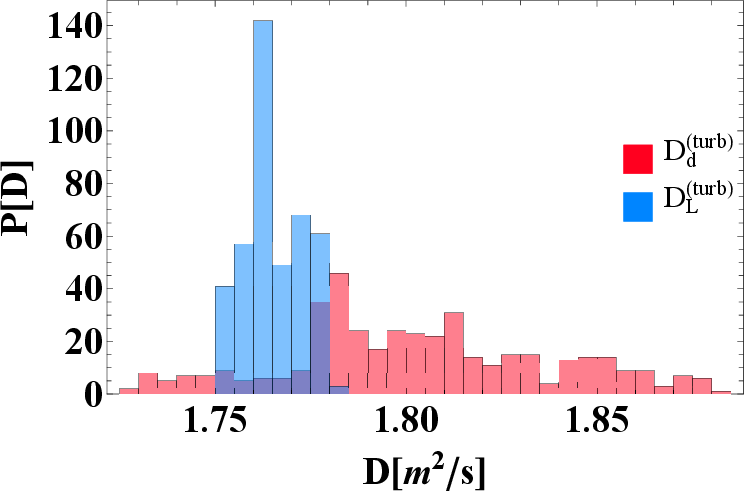}
	}
	\caption{Distribution of $D(t)$ values in the saturated region $t > 50$, computed using $D_d$ (blue) and $D_L$ (red), for the quiescent (a) and turbulent (b) regimes.}
\end{figure}

According to Fig.~\ref{fig_9}, the Lagrangian method of computing diffusion seems to provide a much smoother time-profile that saturates at the same asymptotic values as the differential method $D_d$. This smoothness makes $D_L$ an attractive alternative of computation with much less numerical noise to computing effort ratio. This is further supported by the histograms in Figs.~\ref{fig_10_a}–\ref{fig_10_b}, that show the distributions of saturated ($t > 40$) diffusion values obtained via $D_d$ and $D_L$. 

The explanation for this difference is two-fold. First, the numerical fluctuations in the velocity correlation function are smoothed through time integration in the Lagrangian method, whereas the MSD method amplifies noise due to time differentiation. Second, the uncertainty in MSD scales with the square of the noise amplitude, $\sim A_{\text{noise}}^2$, because it involves squaring particle displacements. In contrast, the uncertainty in the autocorrelation function scales linearly, $\sim A_{\text{noise}}$. Thus, for finite particle ensembles, the Lagrangian method is inherently less sensitive to sampling noise.

\begin{figure}
	\centering
	\includegraphics[width=0.9\linewidth]{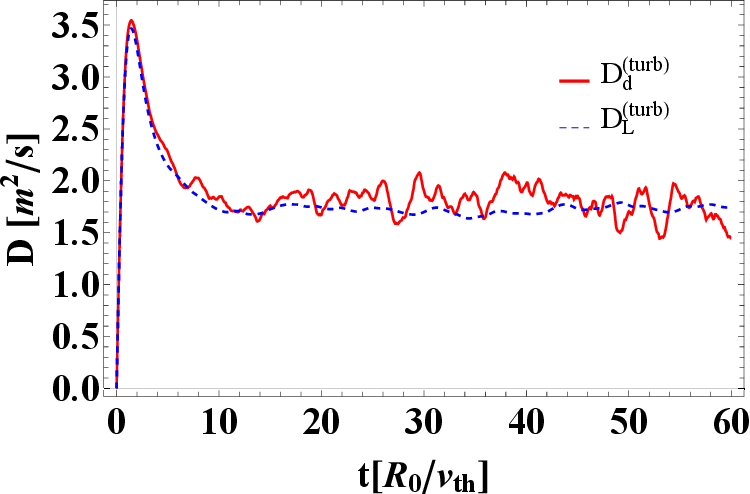}
	\caption{Running diffusion coefficient computed using the MSD and Lagrangian methods for a low-resolution simulation $N_p=60000$.}
	\label{fig_11_a}
\end{figure}

However, this does not mean that $D_L$ is free of numerical fluctuations. In fact, a low-resolution simulations (i.e., small particle numbers $N_p=60000$) is shown in Fig. \ref{fig_11_a} where such oscillations can be seen more clearly. The difference is that they are of lower-frequency than the fluctuations of $D_d$. 

On the other hand, the quantity that has physical relevance is the \emph{asymptotic diffusion coefficient} $D^\infty$ which is calculated in practice \cite{palade2025t3st} (to smooth out numerical noise) via a time-average over the saturated phase:

\begin{eqnarray}\label{eq_3.8.2}
	D_x^\infty = \frac{1}{0.2 t_{max}}\int_{0.8t_{max}}^{t_{max}}D_x(t)dt. 
\end{eqnarray}

\begin{figure}
	\centering
	\subfloat[Asymptotic diffusion coefficient distributions.\label{fig_11_b}]{
		\includegraphics[width=0.9\linewidth]{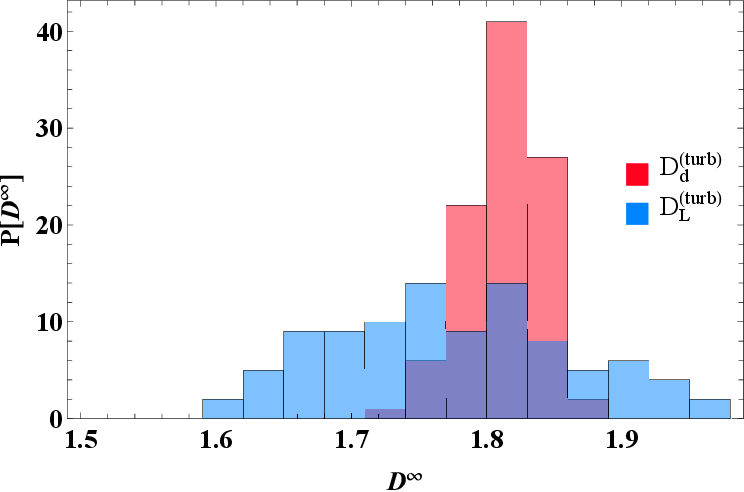}
	}\\
	\subfloat[Time-resolved diffusion with error bars.\label{fig_11_c}]{
		\includegraphics[width=0.9\linewidth]{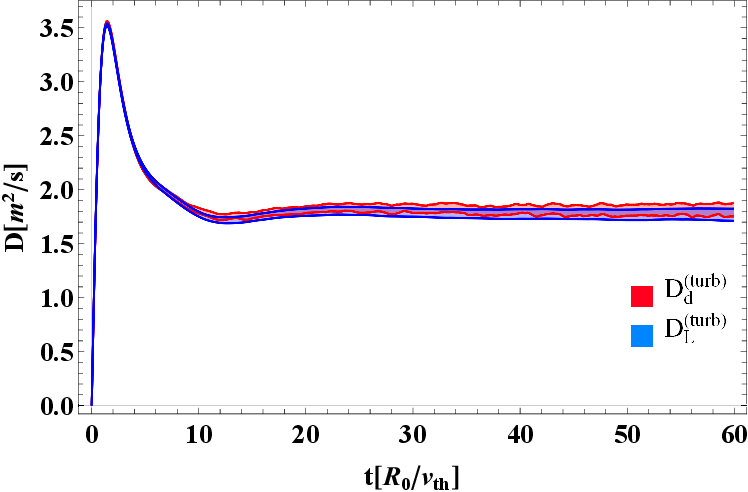}
	}
	\caption{Statistical comparison of the diffusion estimates from the MSD and Lagrangian methods. (a) Distribution of asymptotic diffusion coefficients. (b) Time-resolved diffusion estimates with statistical error bars.}
\end{figure}

The fact that $D_L$ is smoother than $D_d$ across time does not imply necessary that $D_L^\infty$ is more precise than $D_d^\infty$. In fact, Fig. \ref{fig_10_b} already suggests that this value might not match so well. The matter is verified further by performing multiple identical numerical simulations of low resolution and investigate their statistics. This is shown in Fig. \ref{fig_11_b} for the distribution of asymptotic values and in Fig \ref{fig_11_c} for the statistics of running diffusions. The results are rather disappointing. It seems that $D_L^\infty$ are much more prone to numerical fluctuations than $D_d^\infty$ while also providing a smaller average value. Not only that this result is unintuitive but it also raises the question: which is to be trusted, $D_L^\infty$ or $D_d^\infty$?

The answer is that $D_d$ is to be trusted because $D_L$ relies on the stationarity assumption which is only approximately correct. In fact, we can understand that $D_L$, since it works with the correlation $\langle v(0)v(t)\rangle$, is unable to fully capture the growing amplitude of $v(t)$ as shown in Fig. \ref{fig_4}. This is the reason for the consistent underestimation of diffusion.

\begin{figure}
	\centering
	\subfloat[$N_p$ convergence of diffusion coefficients.\label{fig_18_a}]{
		\includegraphics[width=0.9\linewidth]{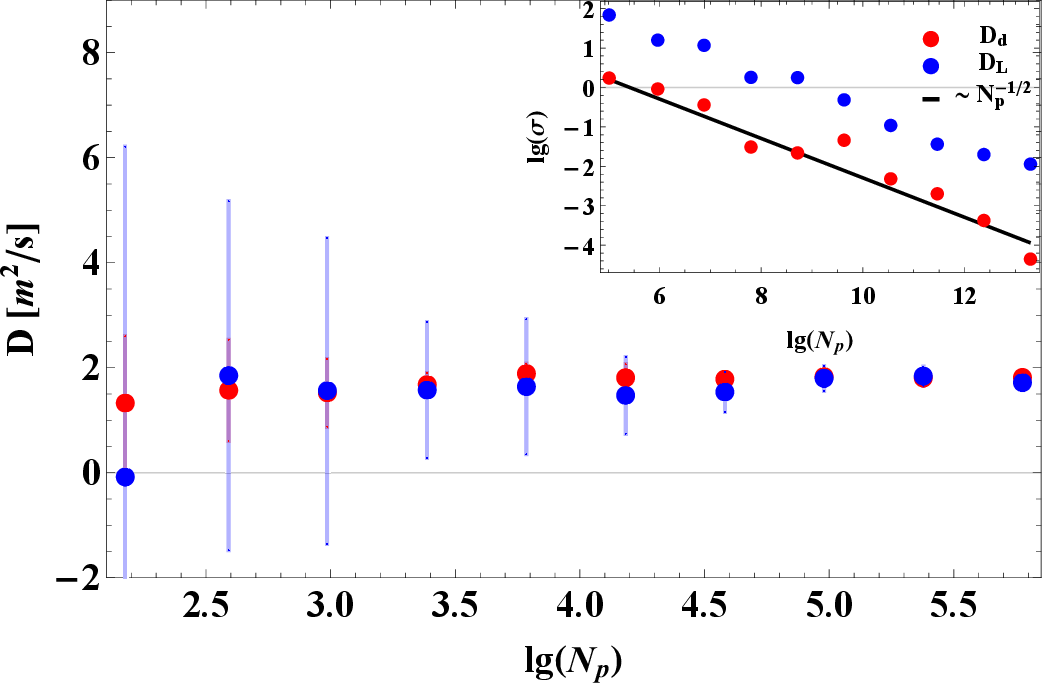}
	}\\
	\subfloat[$N_c$ convergence of diffusion coefficients\label{fig_18_b}]{
		\includegraphics[width=0.9\linewidth]{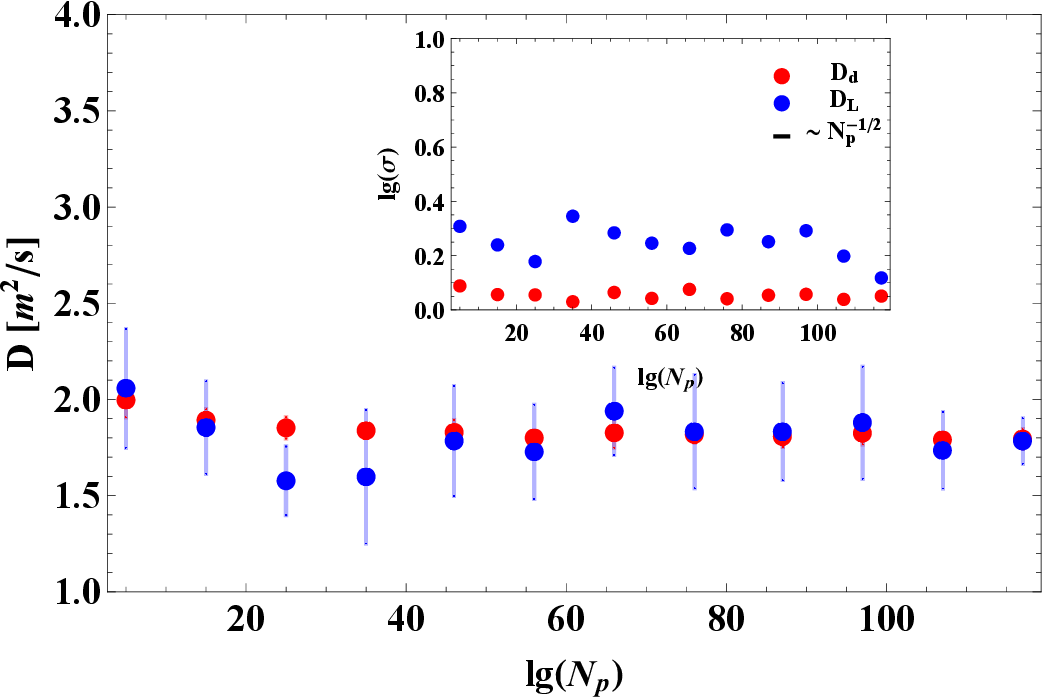}
	}
	\caption{Asymptotic diffusion coefficients with their error bars obtained at different $N_p$ values (a) and $N_c$ values (b) with $D_d$ (red) and $D_L$ (blue). In the inset of each figure, one can see the behavior of the statistical error.}
\end{figure}

In Figs. \ref{fig_18_a} and \ref{fig_18_b} we have investigated the convergence of the statistics of asymptotic diffusion coefficients $D_L^\infty$ and $D_d^\infty$ with the number of test-particles used $N_p$ or partial modes $N_c$. The purpose was to rule out any possible numerical resolution as explanation for the $D_L - D_d$ discrepancies.

\subsection{Lagrangian statistics of a single quiescent trajectory under turbulent drifts}
\label{Section_3.9}

The observed quasi-stationarity of Lagrangian velocities in the quiescent case is particularly striking, given the explicit spatial inhomogeneity and compressibility of magnetic drifts. This observation led us to hypothesize that the apparent stationarity arises not from the drift field itself, but rather from the statistical properties of the \textit{(initial) phase-space distribution} of particles. In other words, the ergodicity required for stationarity may be effectively induced by the broad sampling of phase space present in the kinetic distribution function (ergodic mixing).

To test this hypothesis, we perform a numerical experiment in which a \textit{single quiescent trajectory} is evolved under an ensemble of statistically independent turbulent field realizations. In contrast to the typical approach—where a large ensemble of particles is used to probe a single or multiple field realizations—this setup isolates the contribution of the field ensemble by keeping particle initial conditions fixed. The particle is initialized with a prescribed energy $E = T_i$, pitch angle $\lambda \in \{0.3, 0.8\}$, and located on the low-field-side equatorial plane. The $\lambda =0.3$ corresponds to a trapped banana while $\lambda = 0.8$ to a passing trajectory.

\begin{figure}
	\centering
	\subfloat[Particle tracers in poloidal plane at $t=t_{max}$ for $\lambda = 0.3$ (red) and $\lambda = 0.8$ (blue).\label{fig_17_b}]{
		\includegraphics[width=0.9\linewidth]{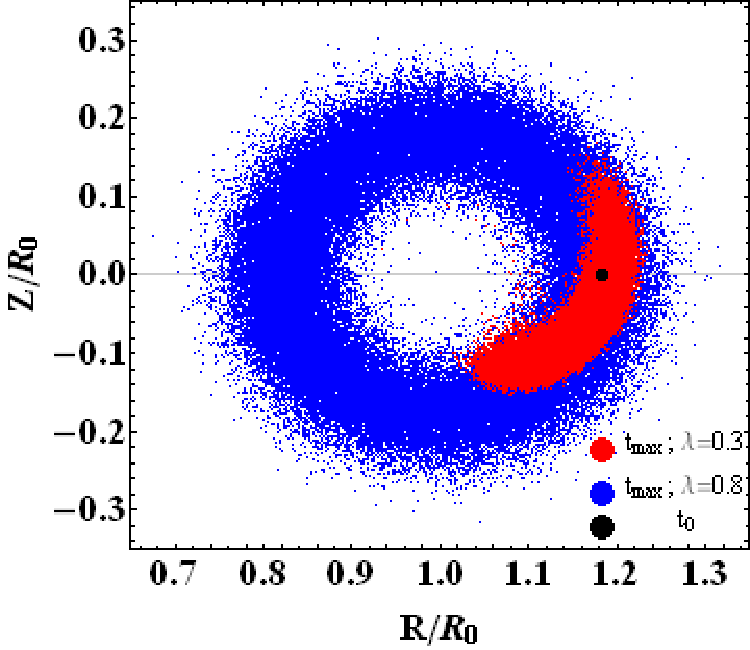}
	}\\
	\subfloat[Histogram of radial positions at $t=t_{max}$ for $\lambda = 0.3$ (red) and $\lambda = 0.8$ (blue).\label{fig_17_c}]{
		\includegraphics[width=0.9\linewidth]{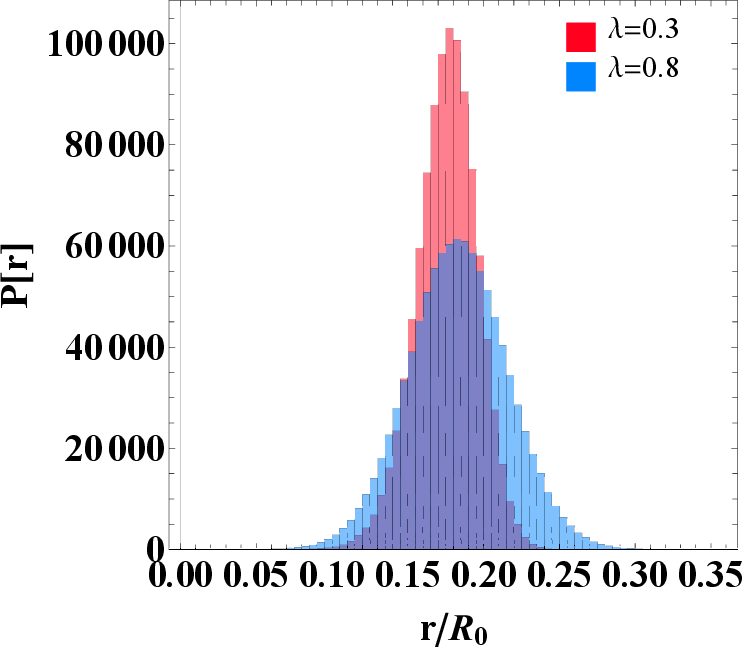}}
	\caption{Distribution of particles at the end of the simulation time for $\lambda = 0.3$ (red) and $\lambda = 0.8$ (blue) in poloidal projection (a) or histogram of radial positions (b).}
\end{figure}

The final distributions of particle trajectories within the statistical ensemble of turbulent fields for the $\lambda = 0.3$ (red) and $\lambda = 0.8$ (blue) pitches is plotted in Figs. \ref{fig_17_b} (poloidal projection)  and \ref{fig_17_c} (radial positions). One can see how passing particles reach until the end of the simulation a quite homogeneous and wide distribution, in contrast to the banana particles that seem to have quite similar trajectories up to longer times. 

\begin{figure}
	\centering
	\includegraphics[width=0.9\linewidth]{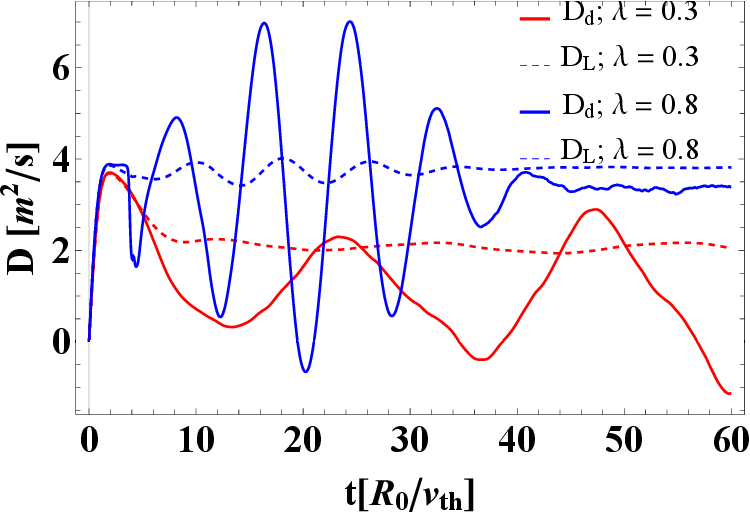}
	\caption{Comparison between running radial diffusions computed with $D_L$ (dashed lines) and $D_d$ (filled lines) for banana (red) and passing trajectory (blue).}	\label{fig_17_a}
\end{figure}

If the particle distribution were solely responsible for the emergence of stationarity, then a single particle would not exhibit stationary velocity statistics, since it cannot sample the full phase space. Indeed, this is confirmed by our results (see Fig. \ref{fig_17_a}): the running diffusion coefficient $D(t)$ computed from the single trajectory under an ensemble of fields shows large fluctuations and it appears to converge at much larger times to a smooth asymptotic value. In contrast, simulations with the full kinetic distribution (i.e., a broad ensemble of initial conditions) consistently produce smooth, saturated profiles. Moreover, the diffusions evaluated with the Green-Kubo relation $D_L$ is unable to follow the MSD $D_d$ (Fig. \ref{fig_17_a}) which implies that stationarity is broken for the turbulent dynamics of a single quiescent trajectory.

\begin{figure}
	\centering
	\subfloat[$\lambda = 0.3$. \label{fig_17_f}]{
		\includegraphics[width=0.9\linewidth]{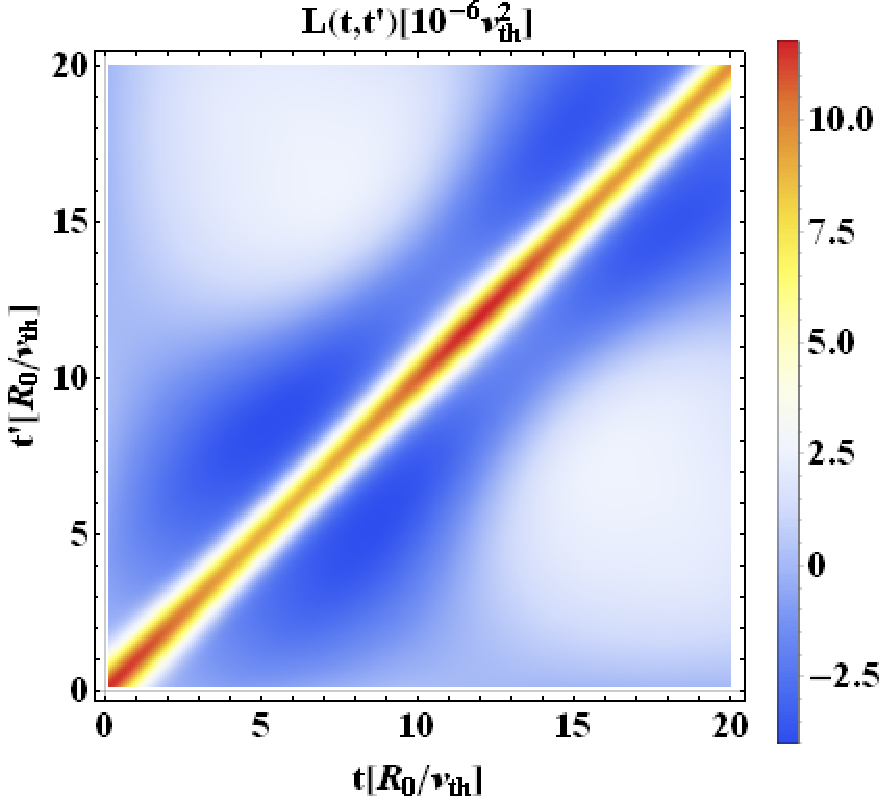}
	}\\
	\subfloat[$\lambda = 0.8$. \label{fig_17_g}]{
		\includegraphics[width=0.9\linewidth]{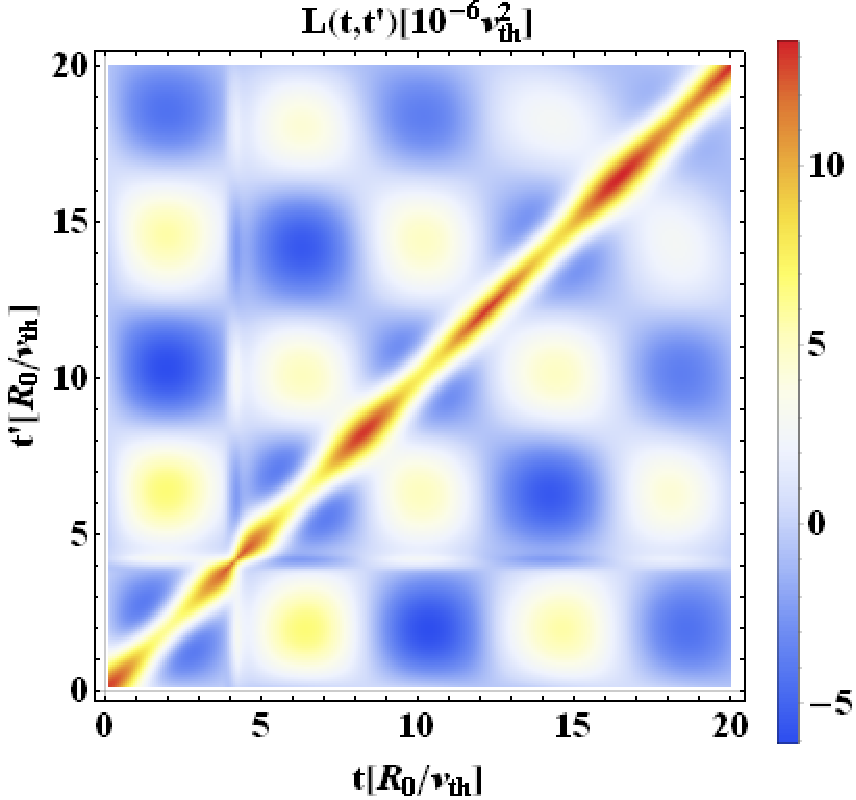}
	}
	\caption{Lagrangian velocity auto-correlation $L(t,t^\prime)$.}
\end{figure}

\begin{figure}
	\centering
	\subfloat[$\lambda = 0.3$.\label{fig_17_e}]{
		\includegraphics[width=0.9\linewidth]{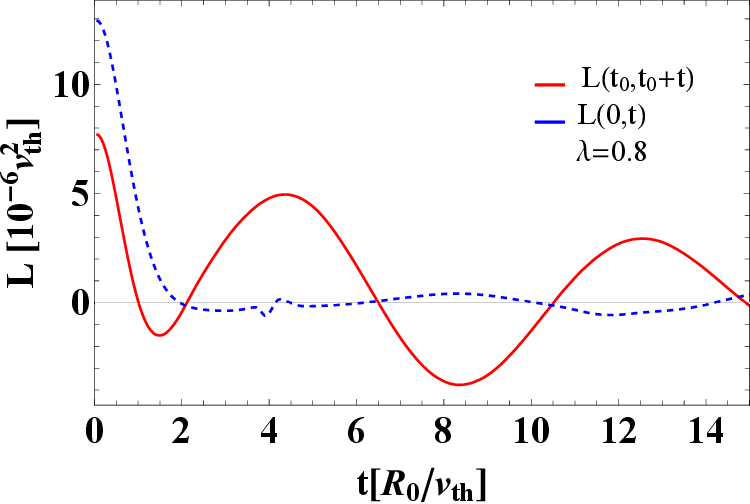}
	}\\
	\subfloat[$\lambda = 0.8$. \label{fig_17_d}]{
		\includegraphics[width=0.9\linewidth]{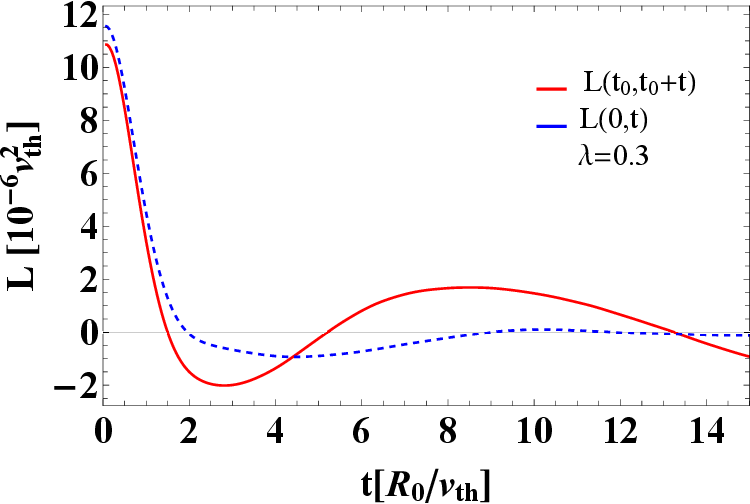}
	}
	\caption{Lagrangian auto-correlation $L(t_0,t_0+t)$ evaluated for $t_0 = 0,15 R_0/v_{th}$ (blue, red lines).}
\end{figure}

This broken stationarity is supported further by plots of the Lagrangian auto-correlation either in 2D such as Fig. \ref{fig_17_f}-\ref{fig_17_g} or in single-time Fig. \ref{fig_17_e}-\ref{fig_17_d}.

These results confirm that the apparent ergodicity and stationarity observed in previous sections are emergent properties of the \textit{joint ensemble of particle states and turbulent fields}—what it is termed a \textit{super-ensemble}. It is the extensiveness of the initial kinetic distribution, spanning a large region of phase space, that enables each particle to sample a quasi-independent portion of the drift field. This effective averaging leads to statistically stationary behavior at the ensemble level, even when the underlying dynamics are spatially inhomogeneous.

This finding emphasizes the importance of correctly modeling the particle distribution function in test-particle simulations of transport. Narrow or non-representative initial distributions may fail to reproduce correct transport properties, not due to inaccuracies in the field model, but due to insufficient sampling of relevant phase-space regions.

\section{Conclusions}
\label{Section_4}

This work presents a comprehensive numerical study of the Lagrangian features of turbulent transport in tokamak plasmas, with a focus on the stationarity, ergodicity, and time-symmetry of test-particle dynamics. Despite the inhomogeneous and compressible nature of the Eulerian drift fields—arising from magnetic curvature and electrostatic turbulence—it is found that Lagrangian velocity statistics are quite robust, exhibiting approximate stationarity and symmetry over time.

Simulations reveal that in the absence of turbulence, magnetic drifts lead to confined, quasi-equilibrium particle distributions with vanishing net transport. In contrast, the introduction of drift-type turbulence results in sustained radial spreading, consistent with a diffusive process. The turbulent regime is characterized by asymptotically Gaussian distributions of both radial displacements and velocities, and a finite, saturated diffusion coefficient. Notably, we identify the presence of a finite radial pinch velocity, attributable to a Turbulent Equipartition (TEP) mechanism linked to magnetic field inhomogeneity.

Through detailed analysis of velocity autocorrelation functions, we confirm that Lagrangian stationarity holds approximately—within a few percent—even in the presence of non-idealities such as field compressibility. Time-symmetry is also found to be preserved in both turbulent and quiescent regimes, as evidenced by symmetric behavior under time-inverted simulations.

Two methods for estimating the diffusion coefficient—the mean square displacement (MSD) method and the Green-Kubo-type correlation method—were tested. While the correlation-based method offers smoother temporal behavior, it may systematically underestimate diffusion due to imperfections of stationarity. Nevertheless, the agreement between methods supports the use of statistical formalisms such as the Decorrelation Trajectory Method (DTM) in fusion plasma modeling.

A particularly important finding is the insensitivity of asymptotic transport properties to the choice of initial particle distribution. Whether particles are initialized from a Maxwell-Boltzmann distribution or a more constrained subset of phase space, the long-time diffusion coefficient remains approximately the same.

In summary, our results provide numerical confirmation that many of the foundational assumptions underlying reduced models of turbulent transport—such as stationarity, ergodicity, and ensemble equivalence—are upheld to good approximation in tokamak-relevant geometries. These findings support the continued use of statistical and semi-analytical tools in fusion transport modeling and may guide future refinements of gyrokinetic theory and test-particle frameworks.

\section*{Acknowledgements}

This work was supported by a grant of the Ministry of Research, Innovation and Digitization, CNCS - UEFISCDI, project number PN-IV-P2-2.1-TE-2023-1102, within PNCDI IV. 


\bibliographystyle{unsrt}
\bibliography{biblio} 

\end{document}